\documentclass[twocolumn, prl, superscriptaddress]{revtex4-1}
\usepackage{amssymb}
\usepackage{amsmath}
\usepackage{epsfig}
\usepackage{graphics, graphicx}
\usepackage{bbold}
\usepackage{psfrag}
\usepackage{mathcomp}
\usepackage{subfigure}
\usepackage{verbatim}
\usepackage{color}
\usepackage[colorlinks,citecolor=blue]{hyperref}
\usepackage{bm}
\def\cp#1{\mathbf{#1}}

\begin{document}

\title{Universal clusters in quasi-two-dimensional ultracold Fermi mixtures}
\author{Ruijin Liu}
\affiliation{Institute of Theoretical Physics, University of Science and Technology Beijing, Beijing 100083, China}
\author{Tingting Shi}
\affiliation{Beijing National Laboratory for Condensed Matter Physics, Institute of Physics, Chinese Academy of Sciences, Beijing, 100190, China}
\author{Matteo Zaccanti}
\affiliation{Istituto Nazionale di Ottica del Consiglio Nazionale delle Ricerche (CNR-INO), 50019 Sesto Fiorentino, Italy}
\affiliation{European Laboratory for Non-Linear Spectroscopy (LENS), Universit$\grave{a}$ di Firenze, 50019 Sesto Fiorentino, Italy}
\author{Xiaoling Cui}
\email{xlcui@iphy.ac.cn}
\affiliation{Beijing National Laboratory for Condensed Matter Physics, Institute of Physics, Chinese Academy of Sciences, Beijing, 100190, China}
\date{\today}

\begin{abstract}
We study universal clusters in quasi-two dimensions (q2D) that consist of  a light (L) atom interacting with two or three heavy (H) identical fermions, forming the trimer or tetramer bound state.
The axial confinement in q2D is shown to lift the three-fold degeneracy of 3D trimer (tetramer) in $p$-wave channel  and uniquely select the ground state  with magnetic angular momentum $|m|=1$ ($m=0$).
By varying the interaction or confinement strength, we explore the dimensional crossover of these clusters from 3D to 2D, characterized by a gradual change of critical H-L mass ratio for their emergence and momentum-space distribution.
Importantly, we find that a finite effective range will {\it not} alter their critical mass ratios  in the weak coupling regime.
There, we establish an effective 2D model to quantitatively reproduce the properties of q2D clusters, and further identify the optimal interaction strengths for their detections in experiments.
Our results suggest a promising prospect for observing universal clusters and associated high-order correlation effects in realistic q2D ultracold Fermi mixtures.
\end{abstract}
\maketitle

The knowledge of few-body bound states is essential for tackling complex many-body problems, since their emergence is usually a precursor of dominant few-body correlation in according many-body systems. In the study of few-body physics, dilute atomic gases have provided an ideal platform and various few-body clusters have been revealed therein~\cite{review_RMP,review_RPP}. Among all of them, the ($1+N$) system that consists of a light (L) atom interacting with $N$ heavy (H) identical fermions emerges as a rare and fascinating case to host {\it universal} cluster bound states~\cite{KM, Blume, Petrov,  Pricoupenko, Parish,Cui, KM_1D,Mehta_1D,Petrov_1D}. Unlike the Efimov states~\cite{Efimov,Castin,Petrov}, these universal clusters are insensitive to short-range details of H-L interactions and are expected to be collisionally stable. Physically, their formation can be attributed to the light-atom-mediated long-range attraction between heavy fermions~\cite{KM,BOA}, and 
only beyond a critical heavy-light mass ratio $\eta_c$ such attraction can overcome the Pauli pressure of heavy fermions to support such universal binding.
Up to date, the critical $\eta_c$ have been successfully extracted in 3D~\cite{KM, Blume, Petrov} and 2D~\cite{Pricoupenko, Parish,Cui} for $N\le 4$ and in 1D~\cite{KM_1D,Mehta_1D,Petrov_1D} for $N\le 5$. The existence of such universal clusters has been shown to intrigue exotic new many-body phases of fermionic matter
~\cite{Parish3, Parish4, Naidon, mass_polaron, QSF, Schmidt}.
A particularly interesting case is 2D, where $\eta_c$ for ($1+N$) clusters are sufficiently low to be accessible by currently available   ultracold Fermi mixtures\cite{footnoot1} such as $^{40}$K-$^{6}$Li~\cite{K_Li1, K_Li2, K_Li3}, $^{161}$Dy-$^{40}$K~\cite{Dy_K1, Dy_K2} and $^{53}$Cr-$^{6}$Li~\cite{Cr_Li, Cr_Li2,Cr_Li3, Cr_Li4}, and where quartet superfluid may emerge~\cite{QSF}:  
a high-order superfluid state beyond the conventional pairing paradigm in two-component fermion systems.

The experimental detection of 2D universal clusters in ultracold gases yet requires two key issues to be resolved. First, in ultracold systems there is no pure 2D, but just quasi-2D (q2D) under strong axial confinement. How clusters behave in q2D is thus an important question for their  practical detection, as previously addressed for Efimov states~\cite{Levinsen, Levinsen2}.  
In particular, a conceptual challenge here is to understand the structural change of universal clusters along the 3D-2D crossover: in 3D, the ground state trimer and tetramer are known to be three-fold degenerate in the $p$-wave channel~\cite{KM, Blume, Petrov}, whereas in 2D such a degeneracy vanishes and the clusters are associated to different angular momenta~\cite{Pricoupenko, Parish,Cui}.  In this context, the q2D study is essentail to bridge distinct few-body physics in different geometries.
The second key question concerns the finite effective range, which has been shown to affect the cluster energy in 3D~\cite{Blume, Petrov}. Here this aspect is crucial for currently available Fermi mixtures~\cite{K_Li1, K_Li2, K_Li3,Dy_K1, Dy_K2,Cr_Li, Cr_Li2,Cr_Li3, Cr_Li4} that generally have narrow Feshbach resonances with large effective range $R^*$ (much larger than Van der Walls length~\cite{Chin}). Testing the robustness of q2D clusters against a large $R^*$ is thus fundamental to their realistic detections.

\begin{figure}[t]
\includegraphics[width=8.5cm]{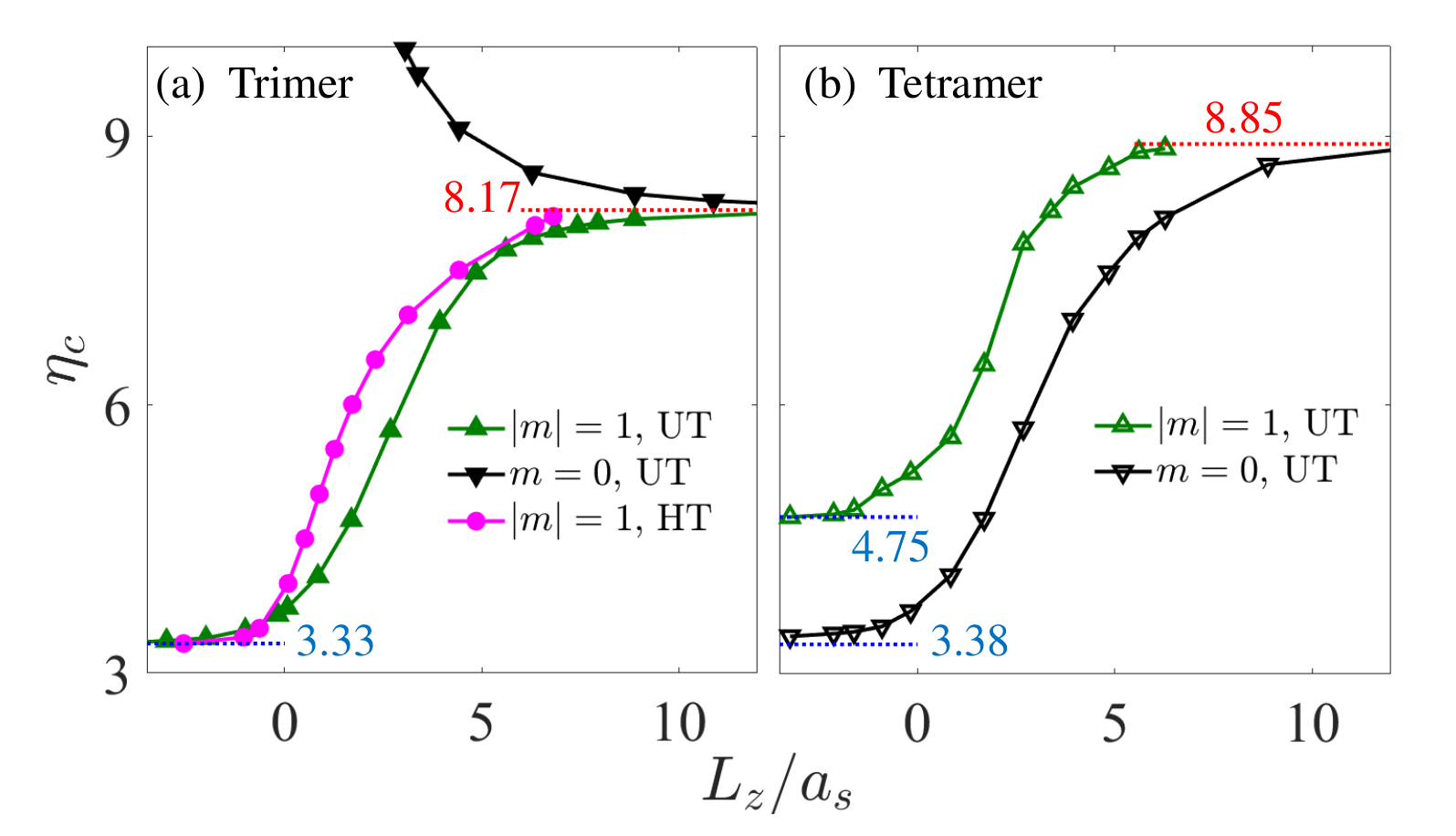}
\caption{(Color Online). Critical heavy-light mass ratios $\eta_c$ for universal trimer (a) and tetramer (b) with different magnetic angular momenta $m=0,\pm 1$ in quasi-2D. In (a) we consider two axial confinements as harmonic trap (HT)~\cite{footnote_HT} and uniform trap (UT), and in (b) we only consider UT. $L_z$ is the axial trap length, and $a_s$ is the 3D s-wave scattering length (assuming $R^*=0$). Horizontal red and blue lines mark $\eta_c$ in our numerics for various clusters in pure 3D and 2D, which were previously studied in Refs.~\cite{KM, Blume, Petrov, Pricoupenko, Parish,Cui}.
}  \label{fig_eta_c}
\end{figure}

\begin{widetext}

\begin{figure}[h]
\includegraphics[width=16cm]{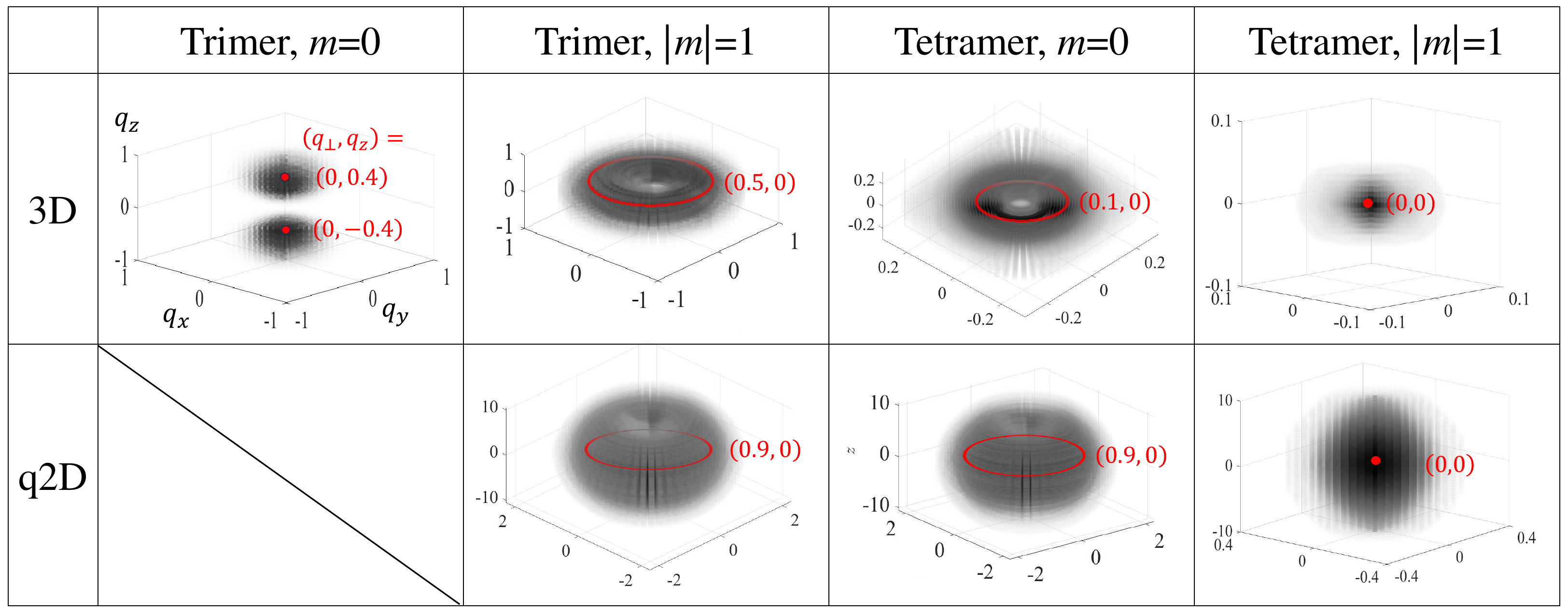}
\caption{(Color Online). Momentum distribution of heavy fermions ($n_h({\cp q})$)    for different clusters at $\eta=9$ in  3D and q2D. The values of $n_h$ increase as the color changes from white to black. Red color denotes the location of maximal $n_h$, with coordinate $(q_{\perp}=\sqrt{q_x^2+q_y^2},q_z)$ shown accordingly. For q2D, we have considered the axial harmonic trap with $L_z/a_s=-1.6$. The tetramer distributions in q2D are from the effective 2D model assuming a frozen motion along $z$ (at the lowest harmonic level), and all other distributions are exact results. The momentum unit is $1/{\bar a}$, with ${\bar a}=(2m_r|\epsilon_2|)^{-1/2}$ the typical dimer size.
}  \label{fig_nk}
\end{figure}

\end{widetext}

\begin{figure}[t]
\includegraphics[width=7cm]{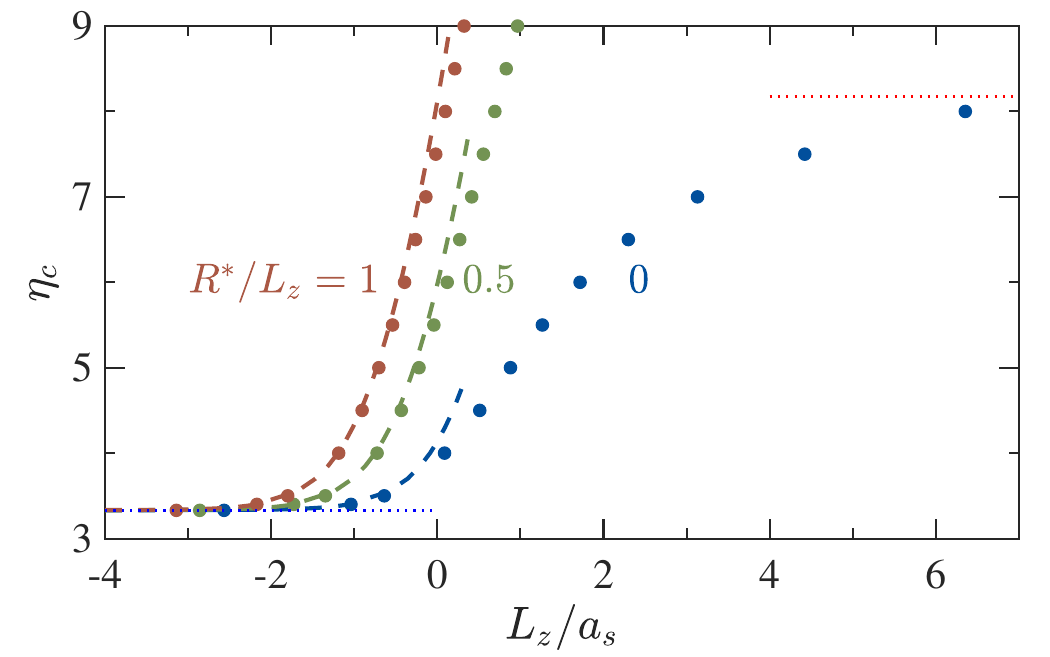}
\caption{(Color Online). Critical mass ratios $\eta_c$ of ground state trimer ($|m|=1$) at different   effective range $R^*/L_z=0,\ 0.5,\ 1$. Here we take the axial harmonic trap. 
Discrete points are exact numerical results and dashed lines are theoretical predictions from the effective 2D model. Horizontal red and blue dotted lines show $\eta_c$ for pure 3D~\cite{KM} and 2D~\cite{Pricoupenko}. }
\label{fig_finite_range}
\end{figure}

\begin{widetext}

\begin{figure}[h]
\includegraphics[width=18cm]{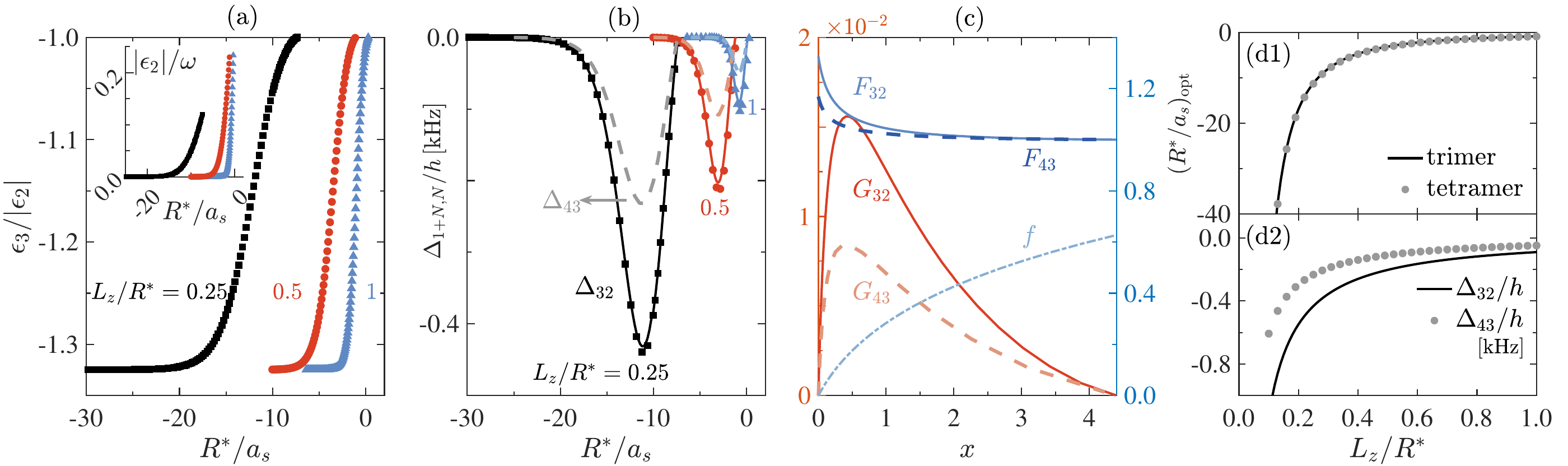} 
\caption{(Color Online). Universal trimer and tetramer in realistic q2D $^6$Li-$^{53}$Cr system ($R^*\simeq6000a_0$)  under an axial harmonic trap.  (a) shows $\epsilon_3/|\epsilon_2|$ as a function of $R^*/a_s$ for different $L_z/R^*$. Inset shows the corresponding $|\epsilon_2|/\omega$. (b)  shows $\Delta_{1+N,N}=\epsilon_{1+N}-\epsilon_{N}$ ($N=2,3$) as functions of $R^*/a_s$ for  different $L_z/R^*$. Discrete points are exact numerical results and continuous lines are predictions from the effective 2D model. (c) shows functions $\{f,\ F_{32},\ F_{43}\}$ and $\{G_{32},\ G_{43}\}$. (d1,d2) show,  respectively, the optimal $R^*/a_s$ and the deepest $\{\Delta_{32},\ \Delta_{43}\}$ computed from the effective 2D model as functions of $L_z/R^*$.
 }
\label{fig_LiCr}
\end{figure}

\end{widetext}

In this work, we positively address these two questions by exactly solving the $(1+N)$ problems in q2D with $N=2$ (trimer) and $N=3$ (tetramer).
By considering either a harmonic or uniform axial confinement, we unveil the general structure of q2D clusters as well as their connection to pure 3D and 2D ones upon changing the confinement or interaction strength. As shown in Fig.~\ref{fig_eta_c} for $R^*=0$, starting from 3D trimer~\cite{KM} and tetramer~\cite{Blume, Petrov}, 
which are both three-fold degenerate with angular momenta $L=1,\ m=0,\pm1$, the application of an axial trap lifts such degeneracy and leads to a splitting of $\eta_c$ between $m=0$ and $m=\pm 1$ channels.
The resulting ground state of q2D trimer  (tetramer) is uniquely associated with $|m|=1$ ($m=0$), well connecting to pure 2D case~\cite{Pricoupenko, Parish,Cui}. Besides $\eta_c$, the crossover of these clusters can also be inferred from their momentum distributions (Fig.~\ref{fig_nk}).
Importantly, a finite $R^*$ is found to {\it hardly} affect $\eta_c$ in the weak coupling regime (Fig.~\ref{fig_finite_range}),  suggesting the robustness of these clusters even under large $R^*$. Focusing on this regime,  we establish an effective 2D model to quantitatively reproduce the properties of these clusters, and further extract the optimal parameters for their experimental detections (Fig.~\ref{fig_LiCr}).  By identifying a feasible route towards the observation of universal clusters, this work paves  the way for future exploration of novel few- and many-body phases in ultracold fermionic matter. 

We start from the Hamiltonian of ($1+N$)-body system in q2D ($\hbar=1$):
\begin{eqnarray}
H&=& \left(-\frac{\nabla^2_{{\cp r}_l}}{2m_{l}} + V_{l}(z_l) \right) +\sum_{i=1}^N \left(-\frac{\nabla^2_{{\cp r}_{h,i}}}{2m_{h}} + V_{h}(z_{h,i}) \right) \nonumber\\
&&\ \ + g\sum_{i=1}^N \delta({\cp r}_l-{\cp r}_{h,i}).
\end{eqnarray}
Here ${\cp r}_l$ and $m_l$ (${\cp r}_{h,i}$ and $m_h$) are the coordinate and mass of the light atom (the $i$-th heavy fermion).  $V_{\sigma}$ ($\sigma=l,h$) is the axial trapping potential, and we will consider two types of $V$: harmonic trap (HT) $V_{\sigma}(z)=m_{\sigma}\omega^2 z_{\sigma}^2/2$ with typical length $L_z=1/\sqrt{2m_r\omega}$ ($m_r=m_hm_l/(m_h+m_l)$ is reduced mass), and  uniform trap (UT) with finite length $L_z$ and periodic boundary condition. 
The bare coupling $g$ is renormalized via $1/g=m_r/(2\pi a_s)-1/V \sum_{\mathbf{Q}}2m_r/\mathbf{Q}^2$, with $a_s$ the 3D s-wave scattering length. A finite effective range ($R^*>0$) can be incorporated in the energy-dependent scattering length~\cite{Chin}
\begin{equation}
a_s^{-1}(E) = a_s^{-1} + R^* (2m_rE),
\end{equation}
with $E$ denoting the energy of two colliding atoms in the center-of-mass (CoM) frame.

We now exactly solve the ($1+N$) problems in q2D. For ($1+1$) dimer, by separating the CoM from relative motions we obtain~\cite{supple} $F(E_2)=0$, where $E_2$ is the energy of relative motion and
\begin{equation}
F(E)=\frac{m_r}{2\pi a_s(E)}-\frac{1}{V} \sum_{\mathbf{Q}}\frac{2m_r}{\mathbf{Q}^2}- \frac{1}{S}\sum_{m,\mathbf{p}}\frac{|\phi_m(0)|^2}{E-\epsilon^z_{m}-\epsilon^{\perp}_{\cp p}}.
\end{equation}
Here $\phi_m(z)$ is the $m$-th eigen-state of relative motion along $z$ with energy $\epsilon^z_{m}$, ${\cp k}$ is the transverse momentum with energy $\epsilon^{\perp}_{\cp k}={\cp k}^2/(2m_r)$ and $S$ is the transverse area. The dimer binding energy is given by $\epsilon_2=E_2-\epsilon^{z}_{m=0}$.

When solving the cluster bound states, we use different methods for different axial confinements. For HT, we solve the ($1+2$) trimer by separating out the CoM motion as done previously~\cite{Levinsen3}. Introducing the relative coordinates ${\cp r}={\cp r}_{h,1}-{\cp r}_l$ and ${\bm{\rho}}={\cp r}_{h,2}-(m_l{\cp r}_l+m_h{\cp r}_{h,1})/(m_l+m_h)$, and imposing the Schr\"{o}dinger equation $H({\cp r},{\bm{\rho}})\Psi_{3}=E_{3}\Psi_{3}$, we arrive at~\cite{supple} 
\begin{eqnarray}
F(E_3-\epsilon^{z;\rho}_{n}-\epsilon^{\perp;\rho}_{\cp k}) f_{n,{\cp k}}=-\frac{1}{S}\sum_{n',{\cp k}'} A_{n{\cp k};n'{\cp k}'} f_{n',{\cp k}'},\label{trimer_eq}
\end{eqnarray}
Here $\epsilon^{\perp;\rho}_{\cp k}={\cp k}^2/(2m_{\rho})$ ($m_{\rho}=m_h(m_h+m_l)/(2m_h+m_l)$), $\epsilon^{z;\rho}_{n}=(n+1/2)\omega$, and $A_{n{\cp k};n'{\cp k}'}$ is the element produced by the Green function when exchanging ${\cp r}_{h,1}\leftrightarrow {\cp r}_{h,2}$.
Physically, $f_{n,{\cp k}}$  is the Fourier transformation of atom-dimer wavefunction.
The trimer binding energy is then given by $\epsilon_3=E_3-\omega$.

For UT, the situation can be greatly simplified since the eigen-states along $z$ are still plane waves. In this case, we are able to solve both ($1+2$) trimer and $(1+3)$ tetramer. Introducing $n$ and ${\cp k}$ as the indices of longitudinal  and transverse momenta, which gives the single-particle energy $\epsilon^{\sigma}_{n{\cp k}}=[n^2(2\pi/L_z)^2+{\cp k}^2]/(2m_{\sigma})$ ($\sigma=l,h$),
we obtain the following coupled equations~\cite{supple}:
\begin{eqnarray}
&&f_{n_2{\cp k}_2...n_N{\cp k}_N}\left( \frac{m_r}{2\pi a_s({\cal E})} - \frac{1}{V} \sum_{\mathbf{Q}}\frac{2m_r}{\mathbf{Q}^2}+ \sum_{n{\cp k}} \frac{(L_zS)^{-1}}{E_{n{\cp k}n_2{\cp k}_2...n_N{\cp k}_N}}\right)
\nonumber\\
&&\  = (L_zS)^{-1}\sum_{{\cp k}} \frac{\sum_{i=2}^N f_{n_2{\cp k}_2...n_i{\cp k}_i...n_N{\cp k}_N}\delta_{nn_i}\delta_{{\cp k}{\cp k}_i}}{E_{n{\cp k}n_2{\cp k}_2...n_N{\cp k}_N}}, \label{FF}
\end{eqnarray}
with $E_{n_1{\cp k}_1n_2{\cp k}_2...n_N{\cp k}_N}=-E_{1+N} +\epsilon^l_{-n_1...-n_N,-{\cp k}_1...-{\cp k}_N}+\sum_{i=1}^N\epsilon^h_{n_i{\cp k}_i}$ and ${\cal E}=E_{1+N} -[(n_2...+n_N)^2(2\pi/L_z)^2+ ({\cp k}_2...+{\cp k}_N)^2]/[2(m_{h}+m_l)]-\sum_{i=2}^N\epsilon^h_{n_i{\cp k}_i}$. Here $E_{1+N}$ directly gives the binding energy $\epsilon_{1+N}$.  After determining $f$-functions from Eqs.~(\ref{trimer_eq},~\ref{FF}), the cluster wavefunctions can also  be obtained~\cite{supple}. 

In our numerics, we have simplified Eqs.~(\ref{trimer_eq},~\ref{FF}) by noting  that the system preserves the total magnetic angular momentum $m_{\rm tot}\equiv m$. It then follows that Eqs.~(\ref{trimer_eq},~\ref{FF}) can be decomposed into  different $m$ sectors, and each $m$-sector  can be solved separately~\cite{supple}. This also allows 
us to sort out the intrinsic relation between q2D clusters and pure 3D/2D ones, as discussed below.

Assuming $R^*=0$, in Fig.~\ref{fig_eta_c} we show the critical mass ratios ($\eta_c=m_h/m_l$) for the emergence of trimer and tetramer in q2D, as given by the conditions $\epsilon_3=\epsilon_2$ and $\epsilon_4=\epsilon_3$, respectively~\cite{supple}.   In the limit  $L_z/a_s\rightarrow+\infty$,  $\eta_c$ approaches $\eta_{c}^{\rm 3D}$ (horizontal red lines) for  3D trimers~\cite{KM} and tetramers~\cite{Petrov, Blume}, which all belong to the $p$-wave channel with three-fold degeneracy ($L=1$ and $|m|=0, 1$). However, as $L_z/a_s$ becomes finite, $\eta_c$ starts to split between $|m|=0$ and $1$, owing to the fact that the rotating symmetry is partly broken by the axial trap and $L$ is no longer a good  number.  The resulting ground state (with a lower $\eta_c$) is solely associated with $|m|=1$ for trimer and  $m=0$ for tetramer. As $L_z/a_s\rightarrow-\infty$, $\eta_c$ of various states continuously decrease and finally saturate at pure 2D values $\eta_{c}^{\rm 2D}$ (horizontal blue lines), as previously studied in Refs.~\cite{Pricoupenko, Cui, Parish}.
The picture in Fig.~\ref{fig_eta_c} reveals the intimate connection of q2D clusters to their 3D/2D counterparts, and this picture is robust against the specific choice of axial confinement, as seen from the results of HT and UT in Fig.~\ref{fig_eta_c}(a).

Beside $\eta_c$, the dimensional crossover of q2D clusters can be reflected in the momentum distribution of heavy fermions, $n_h({\cp q})$.
In Fig.~\ref{fig_nk} we assume $\eta=9$ and show $n_h({\cp q})$ for various clusters along the 3D to 2D crossover.  Notably, different $m$ states feature completely different $n_h({\cp q})$. For 3D trimers, their angular dependence follows $n_h({\cp q})\propto |Y_{1m}(\Omega_{\cp q})|^2$ for each $m$ sector. As a result, the maxima of $n_h({\cp q})$ locate along $z$ axis  for $m=0$, while laying in $xy$ plane for $m=\pm 1$, see Fig.~\ref{fig_nk}. For 3D tetramers, $m=0$ and $m=\pm 1$ states also exhibit distinct $n_h({\cp q})$, whose maxima respectively locate in xy plane with a finite $|{\cp q}_{\perp}|$ and at the origin ${\cp q}=0$.
When crossing over to the 2D regime, the general structures of $n_h({\cp q})$ do not change, except that the distributions along $z$ are frozen at the lowest axial mode~\cite{footnote_nh}.
The unchanged structure also explains why  $m=0$ trimer disappears in 2D, due to the incompatible $z$-distributions between 3D and 2D limits,  as also seen from the tendency  $\eta_c\rightarrow \infty$ in 2D limit in Fig.~\ref{fig_eta_c}. 
In contrast, all other clusters inherit similar distributions from 3D to 2D and thus can all survive in q2D.  
Furthermore, similar to the 2D case~\cite{Cui}, the high-order correlations in q2D clusters can induce crystalline patterns in their  two-body density distributions~\cite{supple}.

We now move to the finite range effect. Similar to 3D clusters\cite{Blume, Petrov},  we find that in q2D a finite $R^*$ generally increases $\eta_c$ and disfavors cluster formation. However, it can {\it hardly} affect $\eta_c$ in the effective 2D regime ($L_z/a_s\ll-1$).  As shown in Fig.~\ref{fig_finite_range} for $|m|= 1$ trimer, as $L_z/a_s\rightarrow -\infty$,  all $\eta_c$ for different $R^*/L_z$   saturate to $\eta_c^{\rm 2D}$, the critical value in pure 2D with $R^*=0$~\cite{Pricoupenko}. We have numerically checked that the saturation $\eta_c\rightarrow \eta_c^{\rm 2D}$ universally applies to clusters under UT~\cite{supple}. This remarkable behavior suggests the detectability of
these clusters in realistic q2D systems even with large $R^*$. For specific $^{40}$K-$^6$Li mixture, we note that the critical  boundaries $\{R^*/L_z, L_z/a_s\}$ in Fig.~\ref{fig_finite_range} agree with those in ~\cite{Levinsen3}.

To understand above behavior, we construct an effective 2D model by utilizing the reduced  scattering length $a_{2D}$ and effective range $R_{2D}$, which enters the 2D scattering amplitude as:
\begin{equation}
T_{2D}(k)=\frac{2\pi}{m_r}\big[-\ln(k^2a_{2D}^2) + i\pi + R_{2D} k^2\big]^{-1}, \label{T}
\end{equation}
here $k$ is the relative collision momentum. 
 For HT, the reduced 2D parameters were derived previously in ~\cite{Petrov2, Kirk07, Parish,Hu2019}, and here we obtain~\cite{supple}
\begin{eqnarray}
a_{2D}&=&\sqrt{\frac{\pi}{0.905}}L_z \exp\left[-\sqrt{\frac{\pi}{2}}(\frac{L_z}{a_s}+\frac{R^*}{2L_z})\right];\nonumber\\
R_{2D}&=&\sqrt{2\pi} R^*L_z+(\ln 2) L_z^2, \label{HT}
\end{eqnarray}
$\{a_{2D}, R_{2D}\}$ for the case of UT are presented in ~\cite{supple}.

Now the universal trend $\eta_c\rightarrow \eta_c^{2D}$ for any $R^*$ in effective 2D regime can be understood. From scaling analysis, in this regime $\eta_c$ only depends on a single parameter $R_{2D}/a_{2D}^2$, which approaches zero as  $L_z/a_s\rightarrow -\infty$ since $a_{2D}\rightarrow\infty$ while $R_{2D}$ is finite. As a result, the finite range plays no effect here, and $\eta^{2D}_c$ is recovered for all $R^*$ (or $R_{2D}$). On the other hand, in the same regime the absolute binding energies of these clusters can be very small, since  all $|\epsilon_{n}|\propto a_{2D}^{-2}\rightarrow 0$. Therefore, it is important to examine how deep the clusters can be bound   under realistic $R^*,\ L_z$ and $a_s$, from which one can identify the optimal parameters for their practical detections.

To do that, we have considered the realistic $^6$Li-$^{53}$Cr system  with a large $R^*\simeq6000a_0$~\cite{Cr_Li,Cr_Li2,Cr_Li3, Cr_Li4} and calculated the absolute binding energies of q2D clusters under HT. Fig.~\ref{fig_LiCr}(a) shows that as decreasing $L_z$, the range of $R^*/a_s$ for  trimer formation ($\epsilon_3/|\epsilon_2|<-1$) expands from a narrow region near resonance to a considerably broader one on $a_s<0$ side. Fig.~\ref{fig_LiCr}(b) further shows $\Delta_{32}\equiv\epsilon_3-\epsilon_2$, which displays a non-monotonic dependence on $R^*/a_s$. For tighter axial trap (smaller $L_z$), the maximum of $|\Delta_{32}|$ becomes larger and its location moves towards weaker coupling, i.e., more negative $R^*/a_s$.

To understand these behaviors, we exploit the effective 2D model to solve the $(1+N)$ problem~\cite{supple}:
\begin{eqnarray}
&&f_{{\cp k}_2...{\cp k}_N}\left( -\sum_{\cp k}\frac{2m_r}{{\cp k}^2+a_{2D}^{-2}}+\frac{Sm_r^2}{\pi}R_{2D}{\cal E}+ \sum_{{\cp k}} \frac{1}{E_{{\cp k}{\cp k}_2...{\cp k}_N}}\right)
\nonumber\\
&&\  =\sum_{{\cp k}} \frac{\sum_{i=2}^N f_{{\cp k}_2...{\cp k}_{i-1}{\cp k}_i{\cp k}_{i+1}...{\cp k}_N}\delta_{{\cp k}{\cp k}_i}}{E_{{\cp k}{\cp k}_2...{\cp k}_N}}, \label{F2}
\end{eqnarray}
with $E_{{\cp k}_1{\cp k}_2...{\cp k}_N}=-\epsilon_{1+N} +\epsilon^l_{-{\cp k}_1...-{\cp k}_N}+\sum_{i=1}^N\epsilon^h_{{\cp k}_i}$ and ${\cal E}=\epsilon_{1+N} -({\cp k}_2...+{\cp k}_N)^2/[2(m_{h}+m_l)]-\sum_{i=2}^N\epsilon^h_{{\cp k}_i}$. 
Utilizing $\{a_{2D},R_{2D}\}$ in Eq.~(\ref{HT}), we find Eq.~(\ref{F2}) yield $\Delta_{32}$ in very good agreement with exact results for $L_z/R^*\lesssim 1$, suggesting the validity of this model even for moderate confinement~\cite{supple}. We further exploit  this model to compute $\Delta_{43}\equiv \epsilon_4-\epsilon_3$, and it shows a similar non-monotonic behavior as $\Delta_{32}$, see Fig.~\ref{fig_LiCr}(b). From the effective 2D model, it is clear that such non-monotonicity is due to the enhanced finite-range effect, i.e., the growing $R_{2D}/a_{2D}^2$ as $1/a_s$ is tuned to strong couplings. 

We now identify the optimal interaction strength for detecting $(1+N)$ clusters and their associated deepest $\Delta_{1+N,N}$. Based on the effective 2D model, we introduce a dimensionless parameter
\begin{equation}
x=R_{2D}/a_{2D}^2 \label{x},
\end{equation}
and express $\epsilon_2$ and $\epsilon_3$ as
\begin{equation}
\epsilon_2=-\frac{1}{m_r R_{2D}}f(x) ;\ \ \ \ \epsilon_3=\epsilon_2 F_{32}(\eta, x).
\end{equation}
Both $f$ and $F_{32}$ are dimensionless functions and can be obtained from Eqs.~(\ref{T}, \ref{F2}).  Then we have $\Delta_{32}=-1/(m_r R_{2D})G_{32}(\eta, x)$, with
\begin{equation}
G_{32}(\eta, x)=(F_{32}(\eta, x)-1)f(x).
\end{equation}
Similarly, for the tetramer we can define $F_{43}(\eta, x)\equiv \epsilon_4/\epsilon_3$ and $\Delta_{43}=-1/(m_r R_{2D})G_{43}(\eta, x)$, with
\begin{equation}
G_{43}(\eta, x)=(F_{43}(\eta, x)-1)F_{32}(\eta,x)f(x).  \label{xx}
\end{equation}

In Fig.~\ref{fig_LiCr}(c) we show the functions $f,\ F_{1+N,N}$ and $G_{1+N,N}$ ($N=2,3$) for Li-Cr system. 
By increasing $x$ from $0$ (2D limit), $f$ monotonically increases while  $F_{1+N,N}$  decreases from a finite value  down to $1$, where $(1+N)$ cluster vanishes. As a result, $G_{1+N,N}$ depends non-monotonically on $x$, and its maximum occurs at $x_m$, which is a universal constant solely relying on $\eta$. Here $x_m$ and its according $G_{1+N,N}$ respectively give the optimal interaction strength and the deepest binding energy for $(1+N)$ cluster. Recalling Eq.~(\ref{HT}), we finally obtain  the optimal $R^*/a_s$ and its associated $\Delta_{1+N,N}$ for given  $L_z/R^*$, see Fig.~\ref{fig_LiCr}(d1,d2). One can see that the optimal condition to detect $\Delta_{1+N,N}$ is the weak coupling regime with strong axial confinement. For instance, at $L_z=0.2R^*$,  the deepest energy detunings for trimer and tetramer can be as large as $|\Delta_{32}|\sim 550$Hz and $|\Delta_{43}|\sim 300$Hz, respectively.
Interestingly, the optimal $R^*/a_s$ that maximize $|\Delta_{32}|$ and $|\Delta_{43}|$ are very close and almost indistinguishable: this can be attributed to the intimate relation between 2D trimer and tetramer~\cite{Cui}, as also inferred by their very close $\eta^{\rm 2D}_c$~\cite{Pricoupenko, Cui}.

In summary, we have revealed the basic structure of universal ($1+N$) clusters 
along the 3D-2D crossover, as classified by different angular momenta and manifested in the evolution of critical mass ratios and momentum distributions. Importantly, it is shown that a finite effective range does {\it not} affect the critical mass ratio for cluster formation in the effective 2D limit, but leads to a non-monotonic binding energy as changing coupling strength. The optimal coupling  and deepest binding energy  have been successfully extracted from an effective 2D model, offering an essential guide for  future detection of these clusters and associated high-order correlation effects in currently available ultracold Fermi mixtures.

We expect that the effective 2D model established here can serve as a convenient tool for tackling general q2D problems with an arbitrary axial confinement. For a general case, our analyses on finite range effect and optimal detection conditions (Eqs.~(\ref{x}-\ref{xx})) are universally applicable once the reduced $\{a_{2D},R_{2D}\}$ are given. Finally, we remark that the disappearance of ground state trimer (tetramer) at critical $\eta_c$ (or at critical $R^*/L_z$ and $L_z/a_s$ for given $\eta$) corresponds to the cluster state hitting the atom-dimer (atom-trimer) scattering threshold. Beyond such a critical boundary, the trimer (tetramer) turns into a p-wave resonance between one dimer (trimer) and a heavy fermion~\cite{Levinsen3,Grimm2014}. This opens a new avenue to explore p-wave physics in effective Bose-Fermi or Fermi-Fermi mixtures.

\bigskip

{\it Acknowledgements.} The work is supported by the National Natural Science Foundation of China (12074419, 12134015), the Strategic Priority Research Program of Chinese Academy of Sciences (XDB33000000). T. S. acknowledges support from the Postdoctoral Fellowship Program of CPSF (No. GZC20232945). R.L. acknowledges support from the National Natural Science Foundation of China (No. 12404316), the Fundamental Research Funds for the Central Universities (No. FRF-TP-24-040A) and the 2023 Fund for Fostering Young Scholars of the School of Mathematics and Physics, USTB (No. FRF-BR-23-01B).

\clearpage

\onecolumngrid
\vspace*{1cm}
\begin{center}
{\large\bfseries Supplementary Materials}
\end{center}
\setcounter{figure}{0}
\setcounter{equation}{0}
\renewcommand{\figurename}{Fig.}
\renewcommand{\thefigure}{S\arabic{figure}}
\renewcommand{\theequation}{S\arabic{equation}}

In this supplementary material, we provide more details on the derivations of few-body equations and reduced 2D scattering parameters, as well as on the properties of universal clusters under an axial uniform trap.

\section*{I.\ \ \ Derivation of few-body equations}

We will first derive the few-body equations of ($1+N$) system  in pure 3D and 2D with a finite effective range, and then proceed with realistic quasi-2D geometry under an axial harmonic or uniform trap.

\subsection*{A. Pure 3D and 2D}

To formally incorporate the finite range effect, we adopt the $D$-dimensional  two-channel model ($D=2,3$):
\begin{equation}
H=\sum_{\mathbf{k}} \left(\epsilon^l_{\mathbf{k}} l_{\mathbf{k}}^{\dagger} l_{\mathbf{k}}+\epsilon^h_{\mathbf{k}} h_{\mathbf{k}}^{\dagger} h_{\mathbf{k}}\right)
+\sum_{\mathbf{k}} \left(\epsilon^d_{\mathbf{k}}+\nu\right) d_{\mathbf{k}}^{\dagger} d_{\mathbf{k}}
+\frac{g}{\sqrt{L^D}} \sum_{\mathbf{k}, \mathbf{Q}} \left(d_{\mathbf{Q}}^{\dagger} h_{\mathbf{k}} l_{\mathbf{Q}-\mathbf{k}}+h.c. \right). \label{eq:H}
\end{equation}
where $\epsilon_{\mathbf{k}}^{l,h}=\mathbf{k}^2/(2m_{l,h})$, $\epsilon_{\mathbf{k}}^{d}=\mathbf{k}^2/(2m_{d})$ ($m_d=m_{l}+m_{h}$) and $L$ is the size of the system along one direction.
The two-body \textit{T}-matrix can be obtained as
\begin{eqnarray}
 \frac{1}{T(E)}=\frac{E-\nu}{g^2}-\frac{1}{L^D}\sum_{\mathbf{k}}\frac{1}{E-\epsilon^l_{\mathbf{k}}
 -\epsilon^h_{\mathbf{k}}},
 \label{T_M_two_channel}
\end{eqnarray}
with $E\equiv q^2/(2m_r)$ the energy of two colliding atoms in the center-of-mass (CoM) frame, and $m_r=m_lm_h/(m_l+m_h)$ the reduced mass.

Comparing with the 3D scattering theory that gives
 \begin{eqnarray}
 \frac{1}{T(E)}=\frac{m_r}{2\pi}\left(\frac{1}{a_s}+R^{*}q^2+iq\right),
 \label{3D_t_matrix}
\end{eqnarray}
one can relate the 3D scattering length $a_s$ and effective range $R^{*}$ to the original parameters ($g,\nu$) in two-channel model via
\begin{eqnarray}
\frac{m_r}{2\pi a_s}=-\frac{\nu}{g^2}+\frac{1}{V}\sum_{\mathbf{k}}\frac{2m_r}{\mathbf{k}^2},\quad\quad
R^{*}=\frac{\pi}{m_r^2g^2}.
\label{RG_3D}
\end{eqnarray}
For 2D system, the 2D scattering theory gives
 \begin{eqnarray}
 \frac{1}{T(E)}=\frac{m_r}{2\pi}\left[-\ln(q^2a^2_{2D})+R_{2D}q^2+i\pi\right]
\label{2D_t_matrix}
\end{eqnarray}
with $a_{2D}$ the 2D scattering length and $R_{2D}$ the effective range.
Similarly, by comparing (\ref{T_M_two_channel}) and (\ref{2D_t_matrix}) we obtain the renormalization relation as:
\begin{eqnarray}
\frac{\nu}{g^2}=\frac{1}{S}\sum_{\mathbf{k}}\frac{2m_r}{a_{2D}^{-2}+\mathbf{k}^2},\quad\quad
R_{2D}=\frac{\pi}{m_r^2g^2}. \label{RG_2D}
\end{eqnarray}

For both 3D and 2D, the two-body binding energy $\epsilon_{2b}\equiv-\kappa^2/(2m_r)$ can be obtained from the singularity of T-matrix, i.e., $T^{-1}(\epsilon_{2b})=0$. Explicitly, in Eqs.~(\ref{3D_t_matrix},\ref{2D_t_matrix}) $q$ has to be replaced by $i\kappa$ according to the analytical continuity.

Now we are ready to solve the ($1+N$) problem by writing down a general ansatz in the CoM frame  as
\begin{eqnarray}
|\Psi\rangle_{1+N} = \sum_{\mathbf{k}_1 \mathbf{k}_2\cdots\mathbf{k}_N } \alpha_{\mathbf{k}_1 \mathbf{k}_2\cdots\mathbf{k}_N} l^{\dag}_{ -{\cp k}_1...-{\cp k}_N} h^{\dag}_{{\cp k}_1}...h^{\dag}_{{\cp k}_N}|0\rangle
+ \sum_{\mathbf{k}_1 \mathbf{k}_2\cdots\mathbf{k}_{N-1} } \beta_{\mathbf{k}_1 \mathbf{k}_2\cdots\mathbf{k}_{N-1}} d^{\dag}_{ -{\cp k}_1...-{\cp k}_{N-1}} h^{\dag}_{{\cp k}_1}...h^{\dag}_{{\cp k}_{N-1}}|0\rangle.
\end{eqnarray}
Imposing the Schr{\"o}dinger equation $H|\Psi\rangle_{1+N}=E_{1+N}|\Psi\rangle_{1+N}$, and defining the function $f_{\mathbf{k}_{2}\cdots\mathbf{k}_{N}}=\frac{g}{\sqrt{L^D}}\sum_{\mathbf{k}_{1}} \alpha_{\mathbf{k}_{1} \mathbf{k}_{2}\cdots\mathbf{k}_{N}}$, we obtain
\begin{eqnarray}
&&f_{{\cp k}_2...{\cp k}_N}\left[\frac{E_{1+N}-\sum_{i=2}^N\epsilon_{\mathbf{k}_{i} }^h-\epsilon_{\mathbf{k}_{2}+\mathbf{k}_{3}+\cdots\mathbf{k}_N }^d-\nu}{g^2}+\frac{1}{L^D}\sum_{\mathbf{k}_{1}} \frac{1}{E_{\mathbf{k}_{1} \mathbf{k}_{2}\cdots\mathbf{k}_{N}}}\right]
 =\sum_{{\cp k}} \frac{\sum_{i=2}^N f_{{\cp k}_2...{\cp k}_{i-1}{\cp k}_i{\cp k}_{i+1}...{\cp k}_N}\delta_{{\cp k}{\cp k}_i}}{E_{{\cp k}{\cp k}_2...{\cp k}_N}}, \label{F2a}
\end{eqnarray}
with $E_{{\cp k}_1{\cp k}_2...{\cp k}_N}=-E_{1+N} +\epsilon^l_{-{\cp k}_1...-{\cp k}_N}+\sum_{i=1}^N\epsilon^h_{{\cp k}_i}$. Eq.~(\ref{F2a}) can be further expressed by effective scattering parameters by utilizing the renormalization relations (\ref{RG_3D}) for 3D and (\ref{RG_2D}) for 2D. Both $f$-function and  $E_{1+N}$ can be solved from (\ref{F2a}).
After determining \textit{f}-functions, we can obtain the original coefficients $\alpha_{\mathbf{k}_1 \mathbf{k}_2\cdots\mathbf{k}_N}$ and $\beta_{\mathbf{k}_1 \mathbf{k}_2\cdots\mathbf{k}_{N-1}}$. Specifically, for the trimer ($N=2$), we have
\begin{eqnarray}
\beta_{\mathbf{k}}&=&\frac{2f_{\mathbf{k}}}{E_3-\epsilon_{\mathbf{k}}^{h}-\epsilon_{\mathbf{k}}^{d}-\nu}
\simeq-\frac{2f_{\mathbf{k}}}{\nu},\nonumber\\
\alpha_{\mathbf{k}_1\mathbf{k}_2}&=&-\frac{g}{2\sqrt{L^D}E_{\mathbf{k}_{1} \mathbf{k}_{2}}}(\beta_{\mathbf{k}_2}-\beta_{\mathbf{k}_1})\simeq \frac{g}{\sqrt{L^D}\nu} \frac{f_{\mathbf{k}_1}-f_{\mathbf{k}_2}}{E_{\mathbf{k}_{1} \mathbf{k}_{2}}}.
\end{eqnarray}
The averaged (one-body) momentum distribution of heavy fermions is then
\begin{equation}
n_h(\mathbf{k})=\sum_{\mathbf{k}_1} \alpha_{\mathbf{k}_1\mathbf{k}}^2+\beta_{\mathbf{k}}^2 \propto \frac{1}{L^D}\sum_{\mathbf{k}_1} \left(\frac{f_{\mathbf{k}_1}-f_{\mathbf{k}}}{E_{\mathbf{k}_{1} \mathbf{k}}}\right)^2 + \frac{4}{g^2}  f_{\mathbf{k}}^2.
\end{equation}
Note that the second term in above equation is proportional to the effective range as defined in Eqs.~(\ref{RG_3D},\ref{RG_2D}).  The two-body momentum distribution reads
\begin{equation}
n_h(\mathbf{k_0},\mathbf{k})= \alpha_{\mathbf{k}_0\mathbf{k}}^2 \propto  \left(\frac{f_{\mathbf{k}_0}-f_{\mathbf{k}}}{E_{\mathbf{k}_{0} \mathbf{k}}}\right)^2
\end{equation}

For the tetramer ($N=3$), we have
\begin{eqnarray}
\beta_{\mathbf{k}_1\mathbf{k}_2}&=&\frac{3f_{\mathbf{k}_1\mathbf{k}_2}}{E_4-\epsilon_{\mathbf{k}_1}^{h}
-\epsilon_{\mathbf{k}_2}^{h}-\epsilon_{\mathbf{k}_1+\mathbf{k}_2}^{d}-\nu}
\simeq-\frac{3f_{\mathbf{k}_1\mathbf{k}_2}}{\nu},\nonumber\\
\alpha_{\mathbf{k}_1\mathbf{k}_2\mathbf{k}_3}&=&-\frac{g}{3\sqrt{L^D}E_{\mathbf{k}_{1} \mathbf{k}_{2}\mathbf{k}_3}}(\beta_{\mathbf{k}_1\mathbf{k}_2}-\beta_{\mathbf{k}_1\mathbf{k}_3}
+\beta_{\mathbf{k}_2\mathbf{k}_3})\simeq \frac{g}{\sqrt{L^D}\nu}\frac{f_{\mathbf{k}_1\mathbf{k}_2}-f_{\mathbf{k}_1\mathbf{k}_3}
+f_{\mathbf{k}_2\mathbf{k}_3}}
{E_{\mathbf{k}_{1} \mathbf{k}_{2}\mathbf{k}_3}}.
\end{eqnarray}
The averaged (one-body) and two-body momentum distribuions of heavy fermions are
\begin{eqnarray}
n_h(\mathbf{k})&=&\sum_{\mathbf{k}_1\mathbf{k}_2} \alpha_{\mathbf{k}_1\mathbf{k}_2\mathbf{k}}^2+\sum_{\mathbf{k}_1}\beta_{\mathbf{k}_1\mathbf{k}}^2
\propto \frac{1}{L^D}\sum_{\mathbf{k}_1\mathbf{k}_2} \left(\frac{f_{\mathbf{k}_1\mathbf{k}_2}-f_{\mathbf{k}_1\mathbf{k}_3}
+f_{\mathbf{k}_2\mathbf{k}_3}}{E_{\mathbf{k}_{1} \mathbf{k}_{2}\mathbf{k}}}  \right)^2 + \frac{9}{g^2}  \sum_{\mathbf{k}_1}f_{\mathbf{k}_1\mathbf{k}}^2;\nonumber\\
n_h(\mathbf{k_0},\mathbf{k})&=& \sum_{\mathbf{k}_1} \alpha_{\mathbf{k}_1\mathbf{k}_0\mathbf{k}}^2 + \beta_{\mathbf{k}_0\mathbf{k}}^2
\propto \frac{1}{L^D}\sum_{\mathbf{k}_1}  \left(  \frac{f_{\mathbf{k}_1\mathbf{k}_0}-f_{\mathbf{k}_1\mathbf{k}}
+f_{\mathbf{k}_0\mathbf{k}}}{E_{\mathbf{k}_{1} \mathbf{k}_{0}\mathbf{k}}} \right)^2 + \frac{9}{g^2}  f_{\mathbf{k}_0\mathbf{k}}^2
\end{eqnarray}
Again the second terms in above equations are proportional to the effective range as defined in Eqs.~(\ref{RG_3D},\ref{RG_2D}).

Similar expressions of $n_h(\mathbf{k})$ and $n_h(\mathbf{k_0},\mathbf{k})$ can also be obtained for q2D case under an axial confinement along $z$, with only difference that the  momentum distribution along $z$ has to be transformed from the eigen-state distribution in this direction. Detailed presentation of these distributions will not be shown here.


\subsection{B. Quasi-2D under an axial confinement}

Here we consider two types of axial confinement: one is uniform trap (UT) and the other is harmonic trap (HT).

\subsubsection{(1). Uniform trap}

For UT with a finite trap size $L_z$ and periodic bound condition, the derivation of few-body equations are similar to pure 3D case, except that the momentum along $z$ is discretized as $k_z=n\frac{2\pi}{L_z}$ ($n=0,\pm1,\pm2,...$) but no longer continuous. The momentum index of a single particle is then $n\mathbf{k}$, with $\mathbf{k}$ the transverse momentum. The final equation for $(1+N)$ system is
\begin{eqnarray}
&&f_{n_2{\cp k}_2...n_N{\cp k}_N}\left[\frac{E_{1+N}-\sum_{i=2}^N\epsilon_{n_i\mathbf{k}_{i} }^h-\epsilon_{n_2+n_3+\cdots n_{N}\mathbf{k}_{2}+\mathbf{k}_{3}+\cdots\mathbf{k}_N }^d-\nu}{g^2}+\frac{1}{L_zS}\sum_{n_1\mathbf{k}_{1}} \frac{1}{E_{n_1\mathbf{k}_{1} n_2\mathbf{k}_{2}\cdots n_N\mathbf{k}_{N}}}\right]
\nonumber\\
&&\  = \frac{1}{L_zS}\sum_{{\cp k}} \frac{\sum_{i=2}^N f_{n_2{\cp k}_2...n_i{\cp k}_i...n_N{\cp k}_N}\delta_{nn_i}\delta_{{\cp k}{\cp k}_i}}{E_{n{\cp k}n_2{\cp k}_2...n_N{\cp k}_N}},
\label{STM_uniform}
\end{eqnarray}
where $E_{n_1{\cp k}_1n_2{\cp k}_2...n_N{\cp k}_N}=-E_{1+N} +\epsilon^l_{-n_{1}...-n_{N};-{\cp k}_1...-{\cp k}_N}+\sum_{i=1}^N\epsilon^h_{n_i{\cp k}_i}$, and $\epsilon_{n\mathbf{k}}^{l,h,d}=\mathbf{k}^2/(2m_{l,h,d})+(\frac{2\pi n}{L_z})^2/(2m_{l,h,d})$. Utilizing the 3D renormalization equation (\ref{RG_3D}), (\ref{STM_uniform}) can be reduced to Eq.~(5) in the main text.

Eq.~(\ref{STM_uniform}) can be simplified by noting that the system preserves total magnetic angular momentum $m_{tot}=m$. Physically, this is because the application of axial confinement (along $z$) will not change the rotation symmetry around $z$-axis, and thus $m$ is still a good quantum number.  In this view, the function $f_{n_2{\cp k}_2...n_N{\cp k}_N}$ can be factorized as
\begin{eqnarray}
f_{n_2{\cp k}_2...n_N{\cp k}_N}=\sum_m f^{(m)}_{n_2k_2...n_Nk_N;\theta_{32}...\theta_{N2}}e^{im\theta_2},
\label{f_uniform}
\end{eqnarray}
where $k_i$ and $\theta_i$ are, respectively, the amplitude and angle of transverse momentum ${\cp k}_i$, and $\theta_{ij}=\theta_i-\theta_j$. Substituting Eq.~(\ref{f_uniform}) into Eq.~(\ref{STM_uniform}), one can obtain the decoupled equations for different $m$-sectors. For each given $m$, we have
\begin{eqnarray}
&&f^{(m)}_{n_2k_2...n_ik_i...n_Nk_N;\theta_{32}...\theta_{i2}...\theta_{N2}}\left[\frac{E_{1+N}-\sum_{i=2}^N\epsilon_{n_i\mathbf{k}_{i} }^h-\epsilon_{n_2+n_3+\cdots n_{N}\mathbf{k}_{2}+\mathbf{k}_{3}+\cdots\mathbf{k}_N }^d-\nu}{g^2}+\frac{1}{L_zS}\sum_{n_1\mathbf{k}_{1}} \frac{1}{E_{n_1\mathbf{k}_{1} n_2\mathbf{k}_{2}\cdots n_N\mathbf{k}_{N}}}\right]
\nonumber\\
& =& -\frac{1}{4\pi^2L_z}\sum_{n_1}\int k_1 d k_1 d\theta_{31}\frac{ f^{(m)}_{n_1k_1n_3k_3...n_Nk_N;\theta_{31}...\theta_{i1}...\theta_{N1}}e^{im(\theta_{32}-\theta_{31})}}{E_{n_1{\cp k}_1n_2{\cp k}_2...n_N{\cp k}_N}}\nonumber\\
&&+\frac{1}{4\pi^2L_z}\sum_{n_1}\int k_1 d k_1 d\theta_{12} \frac{\sum_{i=3}^N f^{(m)}_{n_2k_2...n_ik_i...n_Nk_N;\theta_{32}...\theta_{i2}...\theta_{N2}}\delta_{n_1n_i}\delta_{{\cp k}_1{\cp k}_i}}{E_{n_1{\cp k}_1n_2{\cp k}_2...n_N{\cp k}_N}}
\end{eqnarray}
Specifically, for the trimer and tetramer states we have
\begin{eqnarray}
&&f^{(m)}_{n_2k_{2}}\left[\frac{E_3-\epsilon_{n_2\mathbf{k}_{2} }^h-\epsilon_{n_2,\mathbf{k}_{2} }^d-\nu}{g^2}+\frac{1}{L_zS}\sum_{n_1\mathbf{k}_{1}} \frac{1}{E_{n_1\mathbf{k}_{1} n_2\mathbf{k}_{2}}}\right]=\frac{1}{4\pi^2L_z}\sum_{n_1} \int k_1 d k_1 d\theta_{12}\frac{f^{(m)}_{n_1k_{1}}e^{im\theta_{12}}}{E_{n_1\mathbf{k}_{1} n_2\mathbf{k}_{2}}},
\end{eqnarray}
\begin{eqnarray}
&&f^{(m)}_{n_2k_{2}n_3k_{3};\theta_{32}}\left[\frac{E_4-\epsilon_{n_2\mathbf{k}_{2} }^h-\epsilon_{n_3\mathbf{k}_{3} }^h-\epsilon_{n_2+n_3,\mathbf{k}_{2}+\mathbf{k}_{3} }^d-\nu}{g^2}+\frac{1}{L_zS}\sum_{n_1\mathbf{k}_{1}} \frac{1}{E_{n_1\mathbf{k}_{1} n_2\mathbf{k}_{2}n_3\mathbf{k}_{3}}}\right]\nonumber\\
&&\  =-\frac{1}{4\pi^2L_z}\sum_{n_1} \int k_1 d k_1 d\theta_{31} \frac{f^{(m)}_{n_1k_{1}n_3k_{3};\theta_{31}}e^{im(\theta_{32}-\theta_{31})}}{E_{n_1\mathbf{k}_{1} n_2\mathbf{k}_{2}n_3\mathbf{k}_{3}}}+\frac{1}{4\pi^2L_z}\sum_{n_1} \int k_1 d k_1 d\theta_{12} \frac{f^{(m)}_{n_2k_{2}n_1k_{1};\theta_{12}}}{E_{n_1\mathbf{k}_{1} n_2\mathbf{k}_{2}n_3\mathbf{k}_{3}}}
\end{eqnarray}
Based on these equations, we have numerically solved the q2D trimer ($N=2$) and tetramer ($N=3$) bound states with different $m=0,\pm 1$. 

In principle, the coupled integral equations can be transformed as matrix equations by discretizing the momentum amplitude $k$ and the angle $\theta$. The energy and wavefunction can then be obtained by directly solving the matrix equation. However, the numerical cost for the q2D tetramer in a uniform trap can be very large, especially for the case approaching the 3D limit, where the cutoff of axial quantum number $n$ should be large enough. To overcome this, we treat $k_z=2\pi n/L_z$ as continuous variables for $|n|>9$ when we approach the 3D limit. Moreover,
instead of solving a large matrix equation, we obtain the energy and wavefunction from the coupled equations by using an iterative method. 
We have confirmed the convergence of cluster energy and wavefunction by using different cutoffs and discretization schemes.

\subsubsection{(2). Harmonic trap}

For HT, above treatment is no longer valid because the single-particle states along $z$ are no longer plane-waves. 
In this case, we have solved the $(1+2)$ trimer by separating out the CoM motion and only working  with the relative motions:
\begin{equation}
{\cp r}={\cp r}_{h,1}-{\cp r}_l,\ \ \ \  \bm{\rho}={\cp r}_{h,2}-\frac{m_l{\cp r}_l+m_h{\cp r}_{h,1}}{m_l+m_h}, \label{def}
\end{equation}
where ${\cp r}_{h,1}, {\cp r}_{h,2}$ are the coordinates of two heavy fermions and ${\cp r}_{l}$ is the coordinate of light atom.  Therefore ${\cp r}$ stands for the relative motion between the light atom and one heavy fermion, and $\bm{\rho}$  stands for the relative motion  between this heavy-light pair and the other heavy fermion. Similarly, one can also define another set of relative motions, $({\cp r}_+, {\bm{\rho}}_+)$ by exchanging ${\cp r}_{h,1}\leftrightarrow {\cp r}_{h,2}$ in (\ref{def}). After separating out the CoM motion, the Hamiltonian can be written as
\begin{equation}
H({\cp r}, {\bm{\rho}})=-\frac{\nabla^2_{\cp r}}{2m_r} + \frac{1}{2}m_r\omega^2 r_z^2- \frac{\nabla^2_{\bm{\rho}}}{2m_{\rho}} +  \frac{1}{2}m_{\rho}\omega^2 \rho_z^2 + U\left(\delta({\cp r})+ \delta({\cp r}_+) \right),
\end{equation}
with $m_r=m_hm_l/(m_h+m_l)$ and $m_{\rho}=m_h(m_h+m_l)/(2m_h+m_l)$ the effective mass of ${\cp r}$- and ${\bm{\rho}}$-motions.

The anti-symmetry of the three-body wavefunction $\Psi_3({\cp r}, {\bm{\rho}})$ can be formulated as
\begin{equation}
\Psi_3({\cp r}, {\bm{\rho}})=-\Psi_3({\cp r}_+, {\bm{\rho}}_+).
\end{equation}
In view of such anti-symmetry, we introduce an auxiliary function $f$ through
\begin{equation}
\hat{U}\Psi_3({\cp r}, {\bm{\rho}})=f({\bm{\rho}})\delta({\cp r}) - f({\bm{\rho}}_+)\delta({\cp r}_+),
\end{equation}
with $\hat{U}$ is the interaction operator. Utilizing the Lippman-Schwinger equation, which is equivalent to the Schr\"{o}dinger equation $H({\cp r},{\bm{\rho}})\Psi_{3}=E_{3}\Psi_{3}$:
\begin{equation}
\Psi_3({\cp r}, {\bm{\rho}})=\int d{\cp r}' d{\bm{\rho}}' \langle {\cp r}, {\bm{\rho}}| G_0| {\cp r}', {\bm{\rho}}'\rangle \left( f({\bm{\rho}}')\delta({\cp r}') - f({\bm{\rho}}'_+)\delta({\cp r}'_+) \right),
\end{equation}
where $G_0$ is the non-interacting Green function, we obtain the self-consistent equation for $f$-function:
\begin{equation}
\frac{1}{U} f({\bm{\rho}})= \int d{\bm{\rho}}' \langle {\cp r}=0, {\bm{\rho}}| G_0| {\cp r}'=0, {\bm{\rho}}'\rangle f({\bm{\rho}}') - \int d{\bm{\rho}}'_+ \langle {\cp r}=0, {\bm{\rho}}| G_0 |{\bm{\rho}}'_+, -\frac{m_h}{m_h+m_l}{\bm{\rho}}'_+\rangle f({\bm{\rho}}'_+). \label{eq_f_rho}
\end{equation}
It is more convenient to work with the Fourier transformation of $f$-function by expanding it as
\begin{equation}
f({\bm{\rho}})=\sum_{n,{\cp k}} f_{n,{\cp k}} \Phi_n(\rho_z) \frac{e^{i{\cp k}\cdot {\bm{\rho}}_{\perp}} }{\sqrt{S}}. \label{f_nk}
\end{equation}
Here ${\cp k}$ is the transverse momentum of ${\bm{\rho}}$-motion with energy $\epsilon^{\perp;\rho}_{\cp k}={\cp k}^2/(2m_{\rho})$, and $\Phi_n({\bm{\rho}})$ is the $n$-th eigen-state along $z$  with energy $\epsilon^{z;\rho}_{n}=(n+1/2)\omega$.
Plugging (\ref{f_nk}) into (\ref{eq_f_rho}), we finally obtain the equations in terms of $\{f_{n,{\cp k}}\}$:
\begin{equation}
\left( \frac{1}{U}- \frac{1}{S}\sum_{m,{\cp p}} \frac{|\phi_m(0)|^2}{E_3-\epsilon^{z;\rho}_{n}-\epsilon^{\perp;\rho}_{\cp k}-\epsilon^z_{m}-\epsilon^{\perp}_{\cp p}} \right)f_{n,{\cp k}}=-\frac{1}{S}\sum_{n',{\cp k}'} A_{n{\cp k};n'{\cp k}'} f_{n',{\cp k}'}, \label{final}
\end{equation}
where $\epsilon^z_{m}=(m+1/2)\omega$ and $\epsilon^{\perp}_{\cp p}={\cp p}^2/(2m_r)$ are the energies for ${\cp r}$-motion, $\phi_m(r_z)$ is the according $m$-th eigen-wavefunction along $z$, and $A_{n{\cp k};n'{\cp k}'}$ reads
\begin{eqnarray}
A_{n{\cp k};n'{\cp k}'}&=&
\int dz \ \Phi_n(-\frac{m_hz}{m_h+m_l})\Phi_{n'}(z) \sum_m \frac{\phi_m(0)\phi_m(z)}{E_3-\epsilon^{z;\rho}_{n}-\epsilon^{\perp;\rho}_{\cp k} -\epsilon^z_{m}-\epsilon^{\perp;\rho}_{\cp K}},
\end{eqnarray}
with ${\cp K}\equiv\frac{m_h}{m_h+m_l} {\cp k}+{\cp k}'$. Physically, $f_{n,{\cp k}}$  is the Fourier transformation of atom-dimer wavefunction. To further incorporate the finite range effect, we replace $1/U$ in (\ref{final}) by
\begin{equation}
\frac{1}{U}\rightarrow \frac{m_r}{2\pi a_s(E)}-\frac{1}{V} \sum_{\mathbf{Q}}\frac{2m_r}{\mathbf{Q}^2}, \label{replace}
\end{equation}
with $E=E_3-\epsilon^{z;\rho}_{n}-\epsilon^{\perp;\rho}_{\cp k}$ the relative energy of two colliding atoms, which  can be inferred from the second term in the bracket of (\ref{final}). We have further confirmed the replacement (\ref{replace}) by resorting to a two-channel model, which will not be shown here in detail. With (\ref{replace}),  Eq.~(\ref{final}) can be well reduced to Eq.(4) in the main text.

 Since the total angular momentum $m_{\rm tot}=m$ is conserved, we can project the equation into each $m$-sector. To do this, we expand $f_{n,{\cp k}}$ as
\begin{eqnarray}
f_{n,{\cp k}}=\sum_m f^{(m)}_{nk}e^{im\theta_k},
\end{eqnarray}
with $k,\theta_k$ the amplitude and angle of ${\cp k}$. Then Eq.~(4) in the main text can be well decoupled between different $m$-sectors. For each given $m$, we have
\begin{eqnarray}
F(E_3-\epsilon^{z;\rho}_{n}-\epsilon^{\perp;\rho}_{\cp k}) f_{nk}^{(m)}=-\frac{1}{4\pi^2}\sum_{n'} \int k' dk' d\theta_{k'k} A_{n{\cp k};n'{\cp k}'} f_{n'k'}^{(m)} e^{im\theta_{k'k}},\label{trimer_eq2}
\end{eqnarray}

In our numerics, we discretize the momentum amplitude $k$ and angle $\theta$ to transform the coupled integral equations into a large matrix equation. In the effective 2D limit, where the trimer binding energy is much smaller than $\omega$, we can just take a small cutoff of $n$ in above equation to achieve the convergence of the results. However, when approach to the 3D limit, more and more harmonic oscillator levels have to be included to ensure the convergence. In practice, in this regime we take the cutoff $n_c$ up to $40$, and then extrapolate the results to $n_c\rightarrow \infty$ using a linear fit in terms of $1/n_c$. This strategy works well to the ground state trimer with $|m|=1$. For the excited $m=0$ trimer, however, it needs much heavier numerics to compute each matrix element accurately. As a result, it is difficult to build up a   large matrix (corresponding to cutoff $n_c\sim 40$) within a reasonable time scale. Since we have not obtained the convergent result of trimer energy, we do not show the result for $m=0$ trimer in the main text.

In principle, one can derive the equations for $(1+3)$ tetramer in a similar way as above. However, due to exceedingly heavy numerics brought by the additional degrees of freedom and more matrix elements associated with harmonic oscillator levels, we are currently unable to achieve the exact solution of q2D tetramer under a harmonic trap.

\section{II.\ \ \ Reduced scattering parameters in effective 2D regime}

In this section, we derive the reduced 2D scattering length $a_{2D}$ and effective range $R_{2D}$ for low-energy scattering in q2D.

Considering the relative motion of two atoms, its non-interacting spectrum can be written as
\begin{equation}
\epsilon_{n{\cp k}}=\epsilon_n+\frac{{\cp k}^2}{2m_r},
\end{equation}
with ${\cp k}$ the transverse momentum and $\epsilon_n$ the energy of $n$-th eigen-state along $z$. For the axial uniform trap, we have $\epsilon_n$ and its according eigen-state as
\begin{equation}
\epsilon_n=\frac{(2\pi n/L_z)^2}{2m_r},\ \ \ \ \ \phi_n(z)=\frac{1}{\sqrt{L_z}}e^{2\pi nz/L_z};
\end{equation}
for the axial harmonic trap, we have
\begin{equation}
\epsilon_n=(n+1/2)\omega,\ \ \ \ \ \ \phi_n(z)=(\sqrt{2\pi}2^n n! L_z)^{-1/2}e^{-z^2/(4L_z^2)}H_n(z/(\sqrt{2}L_z)). \ \ \ \ (L_z\equiv 1/\sqrt{2m_r\omega})
\end{equation}

For two-body scattering at low energy $E=\epsilon_0+\frac{{\cp q}^2}{2m_r}$ (${\cp q}^2L_z^2\ll1$), the relative wavefunction can be written as
\begin{equation}
\Psi({\cp r})=\phi_0(z)J_0(q\rho)+f\langle {\cp r}|G_0| {\cp 0}\rangle, \label{psi}
\end{equation}
here $\rho=\sqrt{x^2+y^2}$, $J_0$ is the Bessel function, $f$ represents the scattering amplitude, and $G_0$ is the non-interacting Green function that can be expanded as
\begin{equation}
\langle {\cp r}|G_0| {\cp 0}\rangle=\frac{1}{S}\sum_{n,{\cp k}} \frac{\phi_n^*(0)\phi_n(z)e^{i{\cp k}\cdot{\bm{\rho}}}}{{\cp q}^2/(2m_r)+\epsilon_0-\epsilon_n-{\cp k}^2/(2m_r)+i0^+}. \label{G}
\end{equation}
In the transverse long-distance limit $\rho\rightarrow \infty$, all $n>0$ terms in above equation decays to zero and the wavefunction (\ref{psi}) can be reduced to
\begin{equation}
\Psi({\cp r})\rightarrow \phi_0(z)\left(J_0(q\rho) - f\frac{im_r}{2}\phi_0^*(0) H_0(q\rho)\right), \label{wf_2D}
\end{equation}
with $H_0$ the Hankel function. Comparing (\ref{wf_2D}) to the 2D scattering wavefunction, we have
\begin{equation}
f\frac{im_r}{2}\phi_0^*(0)=\frac{i\pi}{-\ln(q^2a_{2D}^2)+R_{2D}q^2+i\pi}, \label{f_2d}
\end{equation}
where $a_{2D}$ and $R_{2D}$ are the reduced 2D scattering length and effective range. To obtain these parameters, we have to know the information of $f$. This can be done by matching (\ref{psi}) at short distance  to Bethe-Peierls boundary condition
\begin{equation}
\lim_{{\cp r}\rightarrow 0} \Psi({\cp r}) \rightarrow c\left(\frac{1}{r}-\frac{1}{a_s(E)}\right), \ \ \ \ \ \ \left({\rm here}\ \frac{1}{a_s(E)}=\frac{1}{a_s}+R^*(2m_rE)\right). \label{BP}
\end{equation}
In fact, in the short-distance limit (\ref{G}) can be written as
\begin{equation}
\langle {\cp r}|G_0| {\cp 0}\rangle \rightarrow -\frac{m_r}{2\pi r} +C(E), \label{G_short}
\end{equation}
with
\begin{eqnarray}
C(E)&=&\frac{1}{S}\sum_{n=0,{\cp k}} \frac{|\phi_0(0)|^2}{{\cp q}^2/(2m_r)-{\cp k}^2/(2m_r)+i0^+}+\frac{1}{S}\sum_{n>0,{\cp k}} \frac{|\phi_n(0)|^2}{{\cp q}^2/(2m_r)+\epsilon_0-\epsilon_n-{\cp k}^2/(2m_r)}+\frac{1}{V}\sum_{\cp Q} \frac{2m_r}{{\cp Q}^2} \nonumber\\
&=&\frac{m_r}{2\pi} |\phi_0(0)|^2 [-i\pi+F(q^2L_z^2)]. \label{C_E}
\end{eqnarray}
Here $F$ is a dimensionless function that solely depends on $q^2L_z^2$. Combing Eqs.~(\ref{psi}, \ref{G_short},\ref{BP}), we can relate $f$ to 3D scattering parameters $\{a_s,\ R^*\}$. Further recalling Eq.~(\ref{f_2d}), we obtain the following equation which directly relates $\{a_{2D},\ R_{2D}\}$ to $\{a_s,\ R^*\}$:
\begin{equation}
-\ln(q^2a_{2D}^2)+R_{2D}q^2=\frac{1}{|\phi_0(0)|^2} \frac{1}{a_s(E)} - F(q^2L_z^2).
\end{equation}
For the uniform trap, we have $F(x)\rightarrow \ln(x)-x/12$ in the limit $x\ll1$, and therefore
\begin{eqnarray}
a_{2D}=L_ze^{-L_z/(2a_s)},\quad\quad R_{2D}=R^{*}L_z+\frac{L_z^2}{12}.
\label{reduced_uniform}
\end{eqnarray}
For the harmonic trap, in $x\ll 1$ limit we have $F(x)\rightarrow \ln(\pi x/0.905)-(\ln 2)x$,  and therefore
\begin{eqnarray}
a_{2D}=L_{z} \sqrt{\frac{\pi}{0.905}} e^{-\sqrt{\frac{\pi}{2}} [L_z/a_s+R^{*}/(2L_z)]}, \quad\quad
R_{2D}=\sqrt{2\pi}R^{*}L_z+L_z^2\ln 2.
\label{reduced_harmonic}
\end{eqnarray}

\section{III.\ \ \ Numerical results of universal clusters during dimensional crossover}

In this section, we will present more results on the critical emergence and the momentum distribution of universal clusters along the 3D-2D crossover, and we will also discuss the finite-range effect under an axial uniform trap.

\subsection{A. Emergence of universal clusters in q2D}

\begin{figure}[t]
\includegraphics[width=10cm]{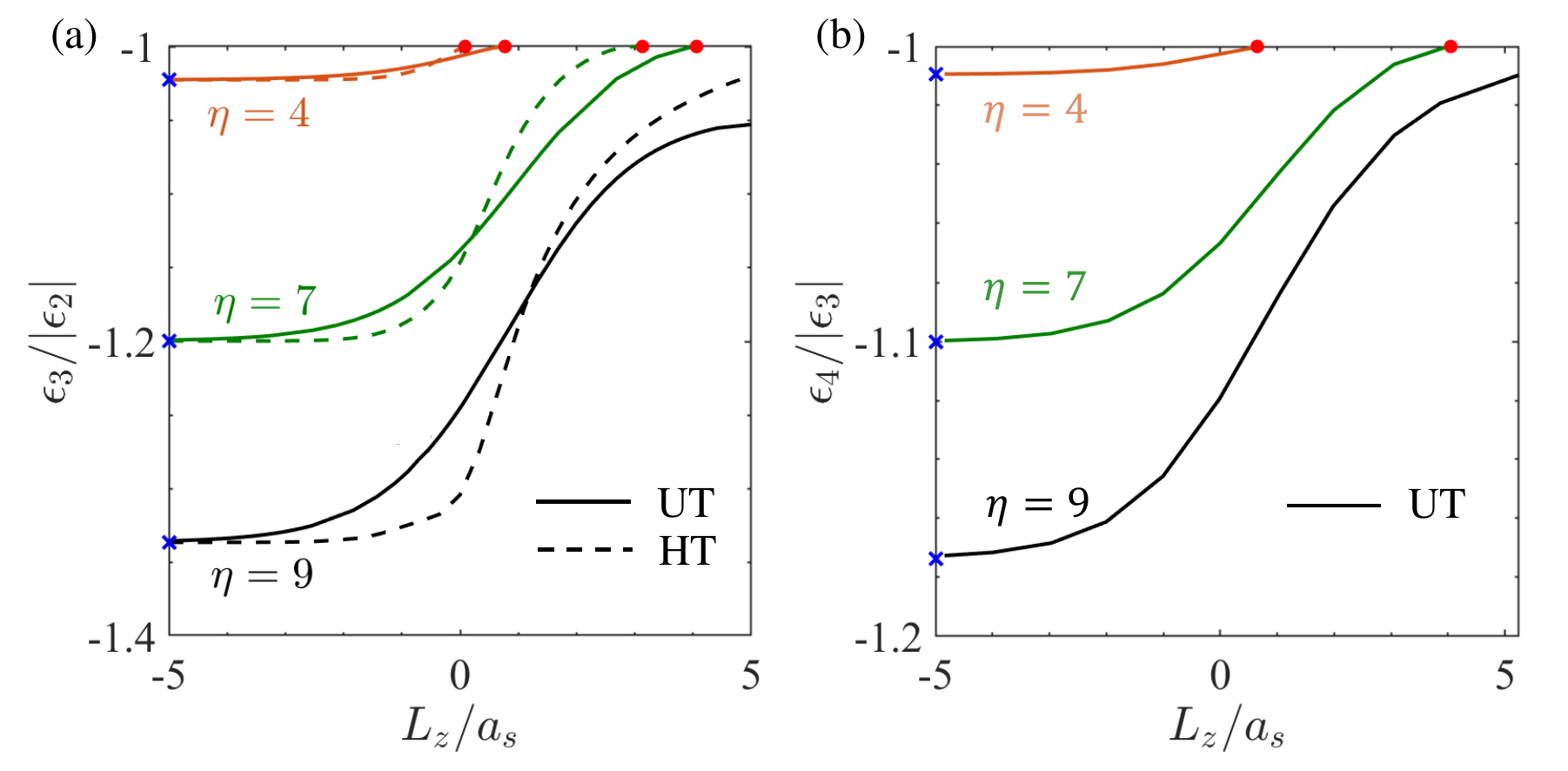}
\caption{(Color Online). $\epsilon_3/|\epsilon_2|$ (a)  and $\epsilon_4/|\epsilon_3|$ (b) as functions of $L_z/a_s$ for different mass ratios $\eta=9, 7, 4$. Here $R^*=0$ is assumed, and $\epsilon_2,\ \epsilon_3$ and $\epsilon_4$ are respectively the binding energies of ground state dimer, trimer and tetramer.
Same as Fig.~1 in the main text,  we have taken both harmonic (HT) and uniform (UT) traps in (a), and just UT in (b). The red points mark the critical $L_z/a_s$ for the emergence of cluster bound states, and blue crossings mark the value of $\epsilon_{1+N}/|\epsilon_N|$ in pure 2D.  }\label{fig_E_ratio}
\end{figure}

To see more clearly the critical emergence of trimer and tetramer in q2D, in Fig.~\ref{fig_E_ratio} we show the binding energy ratios $\epsilon_{1+N}/|\epsilon_N|$ ($N=2,3$) as functions of $L_z/a_s$ for several fixed $\eta$. One can see that for any given $\eta\in(\eta_{c}^{\rm 2D}, \eta_{c}^{\rm 3D})$, the $(1+N)$ cluster bound state will emerge ($\epsilon_{1+N}/|\epsilon_N|\le -1$) as tuning $L_z/a_s$ to certain finite value, which gives the critical boundary shown in Fig.~1 of the main text. In the limit $L_z/a_s\rightarrow-\infty$, $\epsilon_{1+N}/|\epsilon_N|$ saturate at 2D values for each given $\eta(>\eta_{c}^{\rm 2D})$, as marked by crossings in Fig.~\ref{fig_E_ratio}(a,b).

Importantly, Fig.~\ref{fig_E_ratio} shows that the  critical emergence of various ($1+N$) clusters in q2D will depend on both mass ratio ($\eta$) and interaction strength ($L_z/a_s$). This is dramatically different from pure 3D and 2D cases, where the critical emergence of each cluster  only depends on $\eta$ but not the actual scattering length (as long as it is positive).

\subsection{B. Momentum distributions of universal clusters in momentum space}

\begin{figure}[t]
\includegraphics[width=14cm]{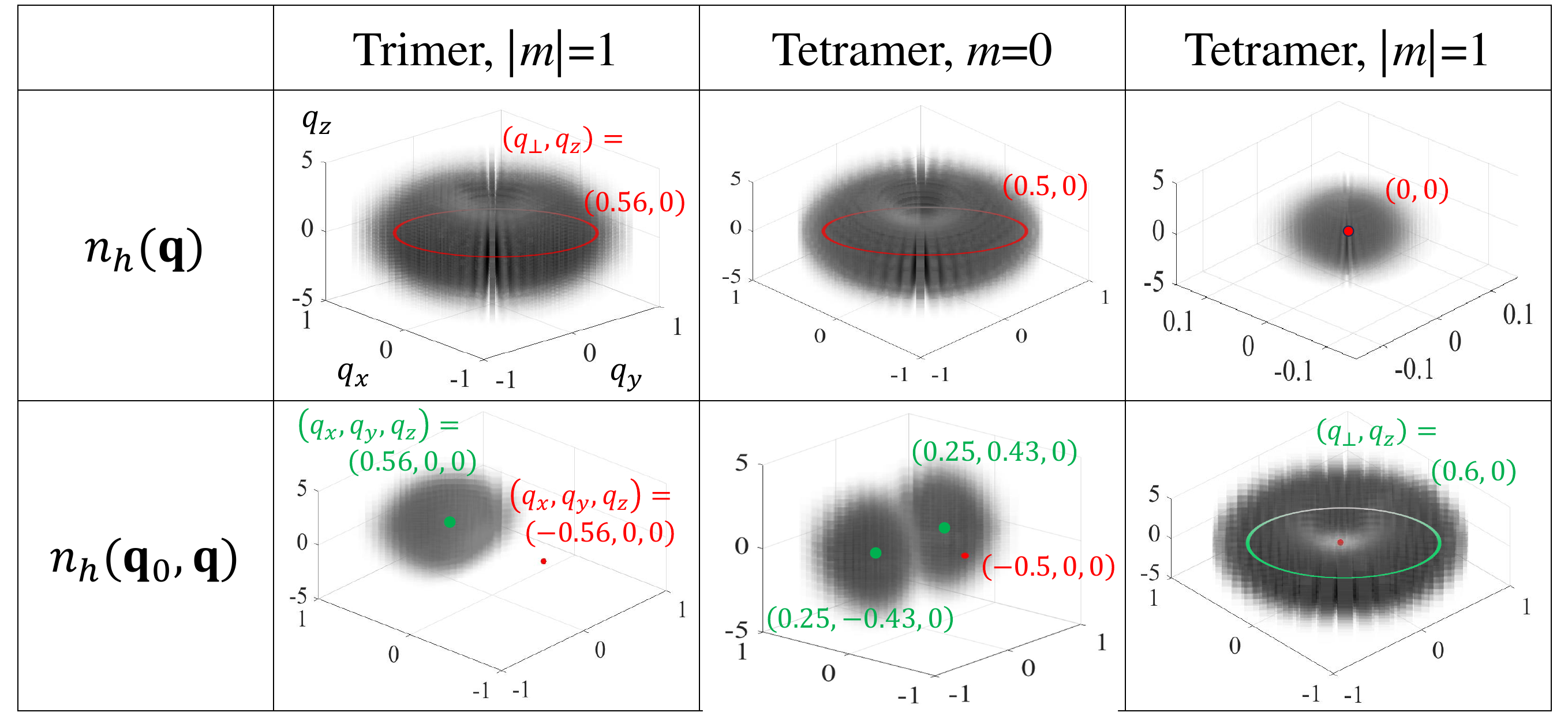}
\caption{(Color Online). One-body ($n_h({\cp q})$) and two-body ($n_h({\cp q}_0,{\cp q})$) density distributions of heavy fermions in momentum space for different clusters in q2D $^6$Li-$^{53}$Cr system under an axial harmonic trap. Here the interaction and confinement parameters are chosen as $R^*/a_s=-11.15$ and $L_z/R^*=0.25$. The values of density distributions increase as the color changes from white to black. For $n_h({\cp q})$ (upper panel), the red point or curve mark its maximum, with coordinate $(q_{\perp}=\sqrt{q_x^2+q_y^2},q_z)$ shown accordingly. For $n_h({\cp q}_0,{\cp q})$, we have chosen ${\cp q}_0$ at one maximum of $n_h({\cp q})$ (see red point), and the green point or curve marks the maximum of $n_h({\cp q}_0,{\cp q})$ with coordinate shown accordingly.
The tetramer distributions are from the effective 2D model assuming a frozen motion along $z$ (at the lowest harmonic level), and the trimer distributions are exact results. The momentum unit is $1/{\bar a}$, with ${\bar a}=(2m_r|\epsilon_2|)^{-1/2}$ the typical dimer size.
}  \label{fig_nk2}
\end{figure}

For an axial harmonic trap, we have computed both the one-body ($n_h({\cp q})$) and two-body ($n_h({\cp q}_0,{\cp q})$)  density distributions of heavy fermions in momentum space for various clusters in q2D.  In Fig.~\ref{fig_nk2}, we show these distributions for a typical interaction strength in the effective 2D regime. In plotting $n_h({\cp q}_0,{\cp q})$, we have chosen ${\cp q}_0$ at a maximum of one-body density distribution. One can see that $n_h({\cp q}_0,{\cp q})$ exhibit spontaneous crystallizations due to intrinsic high-order correlations, as revealed previously for pure 2D clusters. Namely, the ground state ($|m=1|$) trimer exhibits diagonal correlation, and the ground state ($m=0$) tetramer exhibits triangular correlation. In this case, the ground state ($m=0$) tetramer can well distinguish itself from the excited ($|m|=1$) tetramer in the two-body correlations, although they behave quite similar in their one-body distributions (see Fig.~2 in the main text).

\subsection{C. Finite-range effect under an axial uniform trap}

In the main text we have taken the axial harmonic trap to discuss the finite-range effect ($R^*>0$).
As shown in Fig.~3 in the main text, the presence of a finite range will  generally increases the critical mass ratio ($\eta_c$) for the emergence of trimer bound state, but in effective 2D regime it can {\it hardly} affect $\eta_c$, which approaches the pure 2D value ($\eta_c^{2D}$) for any finite $R^*$. In the following  we will study the emergence of universal trimer and tetramer in q2D under an axial uniform trap, and we will show that such finite-range effects equally apply to the uniform trap.

 \begin{figure}[t]
\includegraphics[width=10cm]{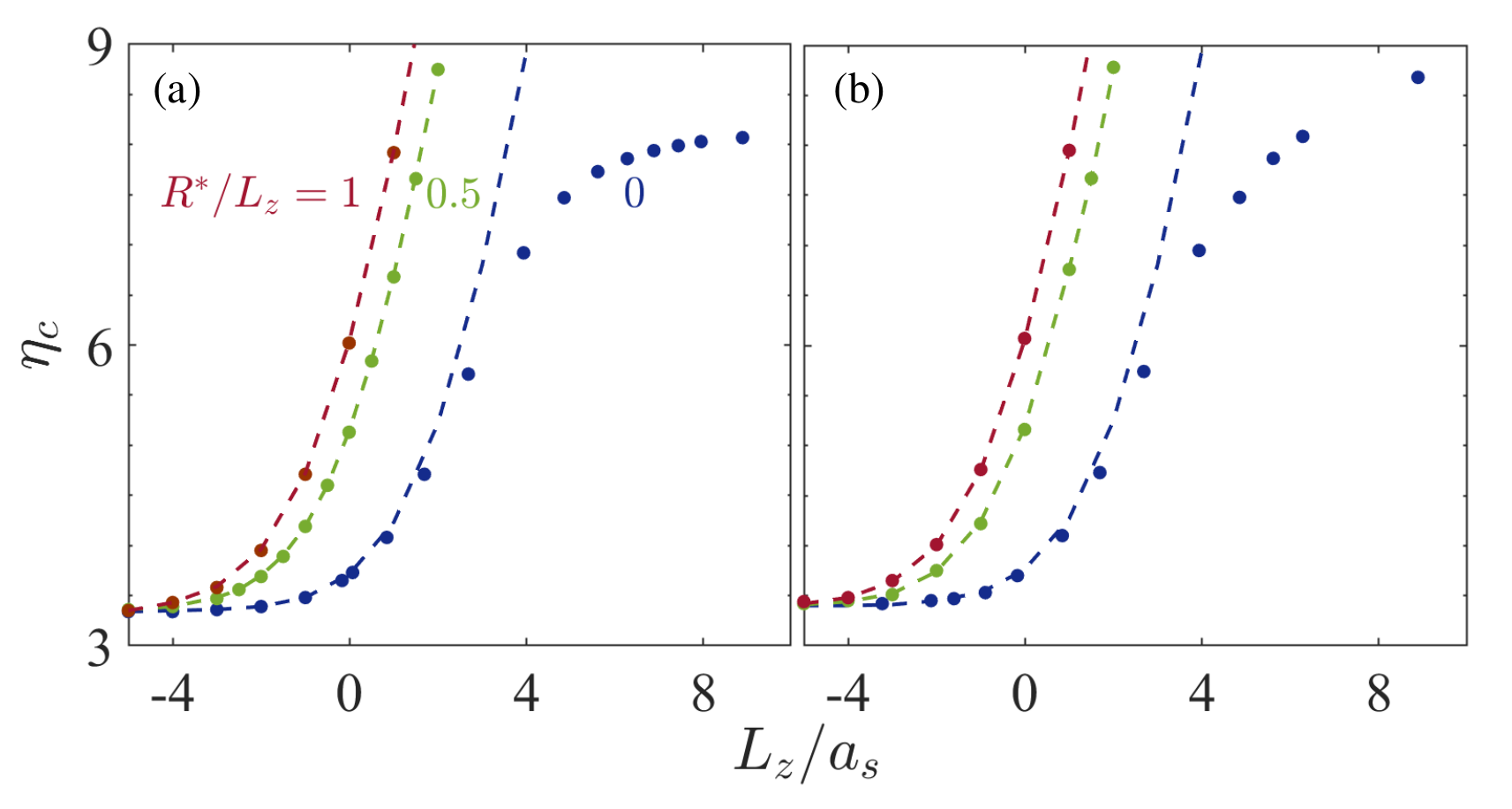}
\caption{(Color Online). Critical mass ratios $\eta_c$ of ground state trimer (a) and tetramer (b) at different   effective range $R^*/L_z=0,\ 0.5,\ 1$ under an axial uniform trap. 
Discrete points are exact results and dashed lines are theoretical predictions based on effective 2D models.}
\label{fig_finite_range_ut}
\end{figure}

Taking an axial uniform trap, in Fig.~\ref{fig_finite_range_ut} we plot out $\eta_c$ as a function of $L_z/a_s$ for the ground state trimer and tetramer with different $R^*$. Similar to the case of harmonic trap,  a finite $R^*$ increases $\eta_c$ for both trimer and tetramer. However, in the effective 2D limit ($L_z/a_s\rightarrow -\infty$), $\eta_c$ for all $R^*$ universally approach the pure 2D values $\eta_c^{2D}$ (3.33 for trimer and 3.38 for tetramer). As analyzed in the main text, this is a direct effect of $R_{2D}/a_{2D}^2\rightarrow 0$ in this effective 2D limit. Here $a_{2D}$ and $R_{2D}$ are respectively the 2D scattering length and effective range under an axial uniform trap, as derived in Eq.~(\ref{reduced_uniform}). Based on such effective 2D model (Eq.~(9) in the main text), we have recalculated $\eta_c$ as a function of $L_z/a_s$ and shown it as dashed line in Fig.~\ref{fig_finite_range_ut}. The predictions fit very well to the numerical results of $\eta_c$ in the effective 2D regime as $L_z/a_s$ gets more negative.


\begin{figure}[t]
\includegraphics[width=10cm]{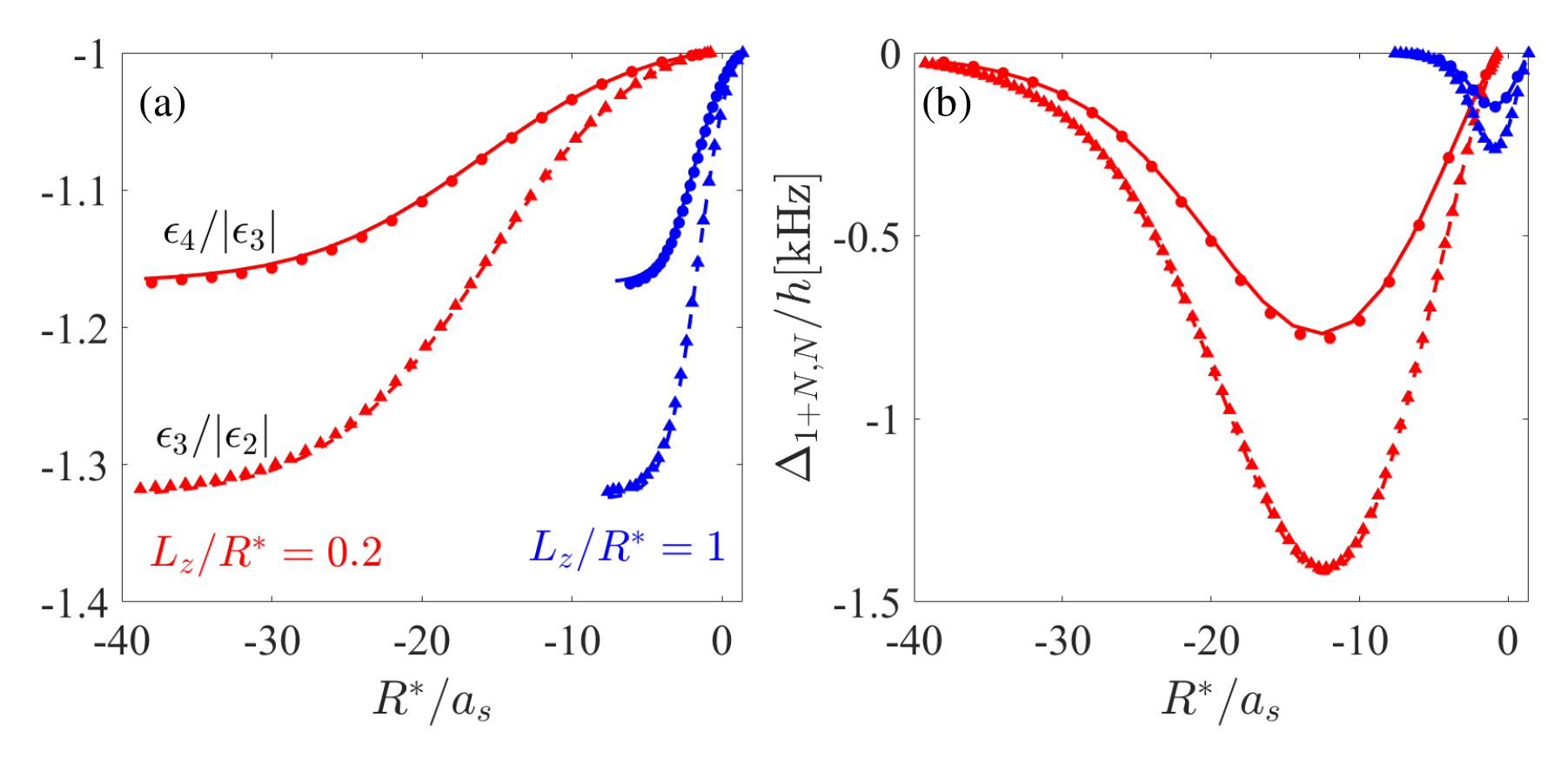}
\caption{(Color Online). Universal trimer and tetramer in q2D $^6$Li-$^{53}$Cr system ($R^*\simeq6000a_0$)  under an axial uniform trap.  (a) shows $\epsilon_3/|\epsilon_2|$ and $\epsilon_4/|\epsilon_3|$ as functions of $R^*/a_s$ at $L_z/R^*=1$ (blue) and $0.2$ (red). (b) shows $\Delta_{32}=\epsilon_3-\epsilon_2$ and $\Delta_{43}=\epsilon_4-\epsilon_3$ as functions of $R^*/a_s$ at $L_z/R^*=1$ (blue) and $0.2$ (red). Discrete points show exact numerical results and continuous lines are theoretical predictions  based on the effective 2D model. }
\label{fig_LiCr_ut}
\end{figure}

The validity of effective 2D model for tight confinement and weak interaction is further confirmed for the uniform trap in Fig.~\ref{fig_LiCr_ut}, where we have taken  the Li-Cr system and computed the energy ratios $\epsilon_{1+N}/|\epsilon_{N}|$ and the detunings $\Delta_{1+N,N}=\epsilon_{1+N}-\epsilon_{N}$ ($N=2,3$) as changing $R^*/a_s$ for different $L_z/R^*=1,\ 0.2$. One can see that the effective 2D model indeed provides  quantitatively good predictions to cluster energies for $L_z/R^*\le1$. In this regime the binding energies of all clusters are much smaller than the axial energy gap, where the effective 2D model can be validated.


\begin{thebibliography}{99}
\bibitem{review_RMP}Chris H. Greene, P. Giannakeas, and J. P${\acute{e}}$rez-R${\acute{i}}$os, \textit{Universal few-body physics and cluster formation}, Rev. Mod. Phys. {\bf 89}, 035006 (2017).

\bibitem{review_RPP}P. Naidon and S. Endo, \textit{Efimov physics: a review}, Rep. Prog. Phys. {\bf 80}, 056001 (2017).


\bibitem{KM}O. I. Kartavtsev and A. V. Malykh, \textit{Low-energy three-body dynamics in binary quantum gases}, J. Phys. B \textbf{40}, 1429 (2007).

\bibitem{Blume}D. Blume, \textit{Universal Four-Body States in Heavy-Light Mixtures with a Positive Scattering Length}, Phys. Rev. Lett. \textbf{109}, 230404 (2012). 
\bibitem{Petrov} B. Bazak and D. S. Petrov, \textit{Five-Body Efimov Effect and Universal Pentamer in Fermionic Mixtures}, Phys. Rev. Lett. \textbf{118}, 083002 (2017).

\bibitem{Pricoupenko}L. Pricoupenko and P. Pedri, \textit{Universal (1+2)-body bound states in planar atomic waveguides}, Phys. Rev. A \textbf{82}, 033625 (2010).
\bibitem{Parish}J. Levinsen and M. M. Parish, \textit{Bound States in a Quasi-Two-Dimensional Fermi Gas}, Phys. Rev. Lett. \textbf{110}, 055304 (2013).
\bibitem{Cui}R. Liu, C. Peng, and X. Cui, \textit{Universal tetramer and pentamer in two-dimensional fermionic mixtures}, Phys. Rev. Lett. \textbf{129}, 073401 (2022).

\bibitem{KM_1D}O. I. Kartavtsev, A. V. Malykh, and S. A. Sofianos, \textit{Bound states and scattering lengths of three two- component particles with zero-range interactions under one-dimensional confinement}, J. Exp. Theor. Phys. \textbf{108}, 365 (2009).
\bibitem{Mehta_1D}N. P. Mehta, \textit{Born-Oppenheimer study of two-component few-particle systems under one-dimensional confinement}, Phys. Rev. A \textbf{89}, 052706 (2014).
\bibitem{Petrov_1D}A. Tononi, J. Givois, and D. S. Petrov, \textit{Binding of heavy fermions by a single light atom in one dimension}, Phys. Rev. A \textbf{106}, L011302 (2022).

\bibitem{Efimov}V. N. Efimov, \textit{Energy levels of three resonantly interacting particles}, Nucl. Phys. \textbf{A210}, 157 (1973).
\bibitem{Castin} Y. Castin, C. Mora, and L. Pricoupenko, \textit{Four-Body Efimov Effect for Three Fermions and a Lighter Particle}, Phys. Rev. Lett. \textbf{105}, 223201 (2010).

\bibitem{BOA}A.C. Fonseca, E.F. Redish, P.E. Shanley, Nucl. Phys. A \textbf{320}, 273(1979).

\bibitem{Parish3}C. J. M. Mathy, M. M. Parish, and D. A. Huse, \textit{Trimers, molecules, and polarons in mass-imbalanced atomic Fermi gases}, Phys. Rev. Lett. \textbf{106}, 166404 (2011).

\bibitem{Parish4}M. M. Parish and J. Levinsen, \textit{Highly polarized Fermi gases in two dimensions}, Phys. Rev. A \textbf{87}, 033616 (2013).

\bibitem{Naidon}S. Endo, A. M. Garc$\acute{i}$a-Garc$\acute{i}$a and  P. Naidon, \textit{Universal clusters as building blocks of stable quantum matter}, Phys. Rev. A {\bf 93}, 053611 (2016).

\bibitem{mass_polaron} R. Liu, C. Peng, X. Cui, \textit{Emergence of Crystalline Few-body Correlations in Mass-imbalanced Fermi Polarons}, Cell Reports Physical Science \textbf{3}, 100993 (2022).

\bibitem{QSF} R. Liu, W. Wang, X. Cui, \textit{Quartet Superfluid in Two-Dimensional Mass-Imbalanced Fermi Mixtures}, Phys. Rev. Lett.  \textbf{131}, 193401 (2023).

\bibitem{Schmidt}R. Li, J. von Milczewski, A. Imamoglu, R. Ołdziejewski and R. Schmidt, \textit{Impurity-induced pairing in two-dimensional Fermi gases}, Phys. Rev. B \textbf{107}, 155135 (2023).

\bibitem{footnoot1} Here we do not consider the Bose-Fermi mixture, since in this system there are additional Efimov-type clusters formed by a fermion and a few bosons.

\bibitem{K_Li1}M. Taglieber, A.-C. Voigt, T. Aoki, T. W. H\"ansch, and K. Dieckmann, \textit{Quantum Degenerate Two-Species Fermi-Fermi Mixture Coexisting with a Bose-Einstein Condensate}, Phys. Rev. Lett. \textbf{100}, 010401 (2008).
\bibitem{K_Li2}E. Wille, F. M. Spiegelhalder, G. Kerner, D. Naik, A. Trenkwalder, G. Hendl, F. Schreck, R. Grimm, T. G. Tiecke, J. T. M. Walraven, S. J. J. M. F. Kokkelmans, E. Tiesinga, and P. S. Julienne, \textit{Exploring an Ultracold Fermi-Fermi Mixture: Interspecies Feshbach Resonances and Scattering Properties of ${}^{6}\textrm{Li}$ and ${}^{40}\textrm{K}$}, Phys. Rev. Lett. \textbf{100}, 053201 (2008).
\bibitem{K_Li3}A.-C. Voigt, M. Taglieber, L. Costa, T. Aoki, W. Wieser, T. W. H\"ansch, and K. Dieckmann, \textit{Ultracold Heteronuclear Fermi-Fermi Molecules}, Phys. Rev. Lett. \textbf{102}, 020405 (2009).

\bibitem{Dy_K1}C. Ravensbergen, V. Corre, E. Soave, M. Kreyer, E. Kirilov, and R. Grimm, \textit{Production of a degenerate Fermi-Fermi mixture of dysprosium and potassium atoms}, Phys. Rev. A \textbf{98}, 063624 (2018).
\bibitem{Dy_K2}C. Ravensbergen, E. Soave, V. Corre, M. Kreyer, B. Huang, E. Kirilov, and R. Grimm, \textit{Resonantly Interacting Fermi-Fermi Mixture of ${}^{161}\textrm{Dy}$ and ${}^{40}\textrm{K}$}, Phys. Rev. Lett. \textbf{124}, 203402 (2020).


\bibitem{Cr_Li}E. Neri, A. Ciamei, C. Simonelli, I. Goti, M. Inguscio, A. Trenkwalder, and M. Zaccanti, \textit{Realization of a cold mixture of fermionic chromium and lithium atoms}, Phys. Rev. A \textbf{101}, 063602 (2020).
\bibitem{Cr_Li2} A.  Ciamei, S. Finelli, A. Trenkwalder, M. Inguscio, A. Simoni and M. Zaccanti, \textit{Exploring ultracold collisions in ${}^{6}\textrm{Li}-{}^{53}\textrm{Cr}$ Fermi mixtures: Feshbach resonances and scattering properties of a novel alkali-transition metal system}, Phys. Rev. Lett. \textbf{129}, 093402 (2022).
\bibitem{Cr_Li3}A. Ciamei, S. Finelli, A. Cosco, M. Inguscio, A. Trenkwalder, and M. Zaccanti, \textit{Double-degenerate Fermi mixtures of ${}^6\textrm{Li}$ and ${}^{53}\textrm{Cr}$ atoms}, Phys. Rev. A \textbf{106}, 053318 (2022).
\bibitem{Cr_Li4}S. Finelli, A. Ciamei, B. Restivo, M. Schemmer, A. Cosco, M. Inguscio, A. Trenkwalder, K. Zaremba-Kopczyk, M. Gronowski, M. Tomza, and M. Zaccanti, \textit{Ultracold LiCr: A New Pathway to Quantum Gases of Paramagnetic Polar Molecules}, PRX Quantum \textbf{5}, 020358 (2024).

\bibitem{Levinsen} J. Levinsen, P. Massignan, and M. M. Parish, \textit{Efimov Trimers under Strong Confinement}, Phys. Rev. X {\bf 4}, 031020 (2014).

\bibitem{Levinsen2}E. K. Laird, T. Kirk, M. M. Parish, J. Levinsen, \textit{Long-lived trimers in a quasi-two-dimensional Fermi system}, Phys. Rev. A {\bf 97}, 042711 (2018).

\bibitem{footnote_HT} $\eta_c$ of $m=0$ trimer under harmonic trap is not shown here given the numerical results are not convergent. 


\bibitem{Chin}C. Chin, R. Grimm, P. Julienne, E. Tiesinga, \textit{Feshbach resonances in ultracold gases}, Rev. Mod. Phys. \textbf{82}, 1225 (2010).

\bibitem{supple}See supplementary material for more details on the derivations of few-body equations and reduced 2D scattering parameters, as well as on the properties of universal clusters under an axial uniform trap.

\bibitem{Levinsen3}  J. Levinsen, T. G. Tiecke, J. T. M. Walraven, D. S. Petrov, \textit{Atom-dimer scattering and long-lived trimers in fermionic mixtures}, Phys. Rev. Lett. \textbf{103}, 153202 (2009).


\bibitem{footnote_nh} For HT, the typical momentum scale of $n_h({\cp q})$ along $z$ is $q_z\sim (m_r\omega)^{1/2}$, more elongated than the dimer scale $\sim (m_r|\epsilon_2|)^{1/2}$ due to $\omega\gg|\epsilon_2|$ in the effective 2D regime.

\bibitem{Petrov2} D. S. Petrov and G. V. Shlyapnikov, \textit{Interatomic collisions in a tightly confined Bose gas}, Phys. Rev. A \textbf{64}, 012706 (2001).

\bibitem{Kirk07} T. Kirk, M. M. Parish, \textit{Three-body correlations in a two-dimensional SU(3) Fermi gas}, Phys. Rev. A \textbf{96}, 053614 (2017).

\bibitem{Hu2019} H. Hu, B. C. Mulkerin, U. Toniolo, L. He, and X.-J. Liu,
\textit{Reduced Quantum Anomaly in a Quasi-Two-Dimensional Fermi Superfluid: Significance of the Confinement-Induced Effective Range of Interactions}, Phys. Rev. Lett. \textbf{122}, 070401 (2019).


%

\bibitem{Grimm2014} M. Jag, M. Zaccanti, M. Cetina, R. S. Lous, F. Schreck, R. Grimm, D. S. Petrov, and J. Levinsen, \textit{Observation of a Strong Atom-Dimer Attraction in a Mass-Imbalanced Fermi-Fermi Mixture}, Phys. Rev. Lett. \textbf{112}, 075302 (2014).

\end{thebibliography}
\end{document}